%% file: tveg.tex
\documentclass{egpubl}
\JournalPaper         % uncomment for final version of Journal Paper
\usepackage[T1]{fontenc}
\usepackage{dfadobe}  

\usepackage{cite}  % comment out for biblatex with backend=biber
\BibtexOrBiblatex
\electronicVersion
\PrintedOrElectronic
\ifpdf \usepackage[pdftex]{graphicx} \pdfcompresslevel=9
\else \usepackage[dvips]{graphicx} \fi

\usepackage{egweblnk}

\usepackage{amsmath}
\usepackage{amssymb}
\usepackage{amsfonts}
\usepackage{graphicx}
\usepackage{subfigure}
\usepackage{color}
\usepackage[dvipsnames]{xcolor}
\usepackage{xspace}
\usepackage{mathtools}

\usepackage[linesnumbered,ruled]{algorithm2e}
\DeclareMathOperator*{\argmin}{arg\!min}
\graphicspath{{./fig/}}

\SetCommentSty{mycommfont}

\newcommand{\LWM}{\textsc{lwm}\xspace}
\newcommand{\TVEG}{\textsc{tveg}\xspace}
\newcommand{\ie}{\textit{i.e.},\xspace}

\newcommand{\minorrevision}[1]{{{#1}}\xspace}
\newcommand{\majorrevision}[1]{{{#1}}\xspace}
\newcommand{\minoredits}[1]{{{#1}}\xspace}
\newcommand{\highlightadd}[1]{{{#1}}\xspace}

\title[Time-varying Extremum Graphs]%
      {Time-varying Extremum Graphs}

\author[S. Das, R. Sridharamurthy \& V. Natarajan]
{\parbox{\textwidth}{\centering
        Somenath Das\orcid{0000-0002-3582-7943},
        Raghavendra Sridharamurthy\orcid{0000-0001-8463-0488}, and
        Vijay Natarajan\orcid{0000-0002-7956-1470}
        }
        \\
{\parbox{\textwidth}{\centering
Department of Computer Science and Automation, Indian Institute of Science, Bangalore, India}
}
}

\begin{document}

% uncomment for using teaser
% \teaser{
%  \includegraphics[width=\linewidth]{eg_new}
%  \centering
%   \caption{New EG Logo}
% \label{fig:teaser}
%}

\maketitle
%-------------------------------------------------------------------------
\begin{abstract}
We introduce time-varying extremum graph (\TVEG), a topological structure to support visualization and analysis of a time-varying scalar field. The extremum graph is a substructure of the Morse-Smale complex. It captures the adjacency relationship between cells in the Morse decomposition of a scalar field. We define the \TVEG as a time-varying extension of the extremum graph and demonstrate how it captures salient feature tracks within a dynamic scalar field. We formulate the construction of the \TVEG as an optimization problem and describe an algorithm for computing the graph. We also demonstrate the capabilities of \TVEG towards identification and exploration of topological events such as deletion, generation, split, and merge within a dynamic scalar field via comprehensive case studies including a viscous fingers and a 3D von K\'arm\'an vortex street dataset.



\end{abstract}  
%-------------------------------------------------------------------------
\input{section1.tex}
\input{section2.tex}

\input{section3.tex}
\input{section4.tex}
\input{section5.tex}

\section*{Acknowledgments}
This work is partially supported by a grant from SERB, Govt. of India (CRG/2021/005278), CV Raman Postdoctoral Fellowship from IISc Bangalore, and MoE Govt. of India.
VN acknowledges support from the Alexander von Humboldt Foundation, and Berlin MATH+ under the Visiting Scholar program. Part of this work was completed when VN was a guest Professor at the Zuse Institute Berlin.

\noindent \textbf{Data Availability.} The data that support the findings of this study are available from the corresponding author upon reasonable request.

\noindent \textbf{Conflict of Interest.} None.

\bibliographystyle{eg-alpha-doi}
\bibliography{tveg}

\input{tveg_app_cgf}
\end{document}

%% file: section1.tex
\section{Introduction}\label{sec_introduction}
The study and development of effective methods for analysis and visualization of time-varying scalar fields continues to be a challenging problem due to the geometric complexity of the data. The study of many scientific processes naturally requires the computation or measurement of scalar fields and their time-varying counterparts. A comprehensive analysis of a time-varying field benefits from a global view as compared to independent analysis of the individual time steps. An animation  helps the user gain a cursory understanding but a feature-directed approach towards visualization and analysis of the field is required for a comprehensive understanding. Efficient representation of features in the data and methods for tracking their evolution are crucial ingredients of such an approach. Data is continuously increasing in size and becoming feature rich, necessitating methods for succinct and abstract representations of the features and their evolution. Topological structures like the persistence diagram, Reeb graph, contour tree, and Morse-Smale (MS) complex were developed to address this challenge. 

\begin{figure}

\centering
 
\subfigure[Scalar field $f$]{\includegraphics[width=0.21\textwidth]{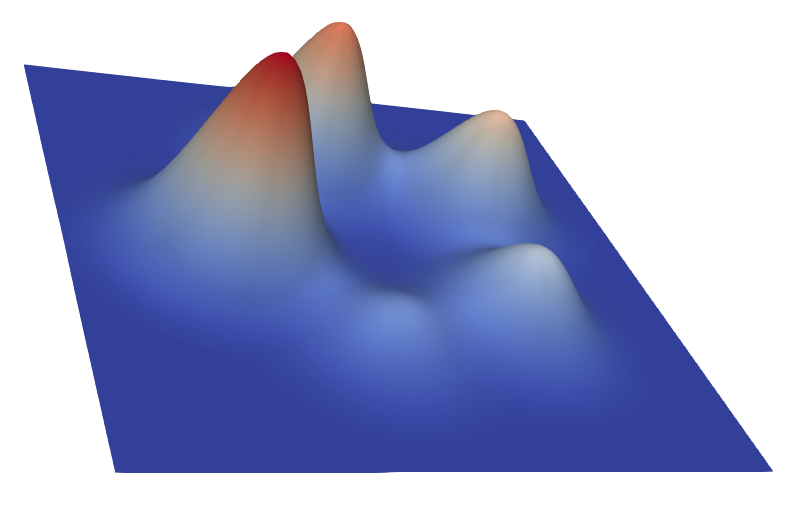}
\label{fig:field}}
~
\subfigure[Critical points]{\includegraphics[width=0.21\textwidth]{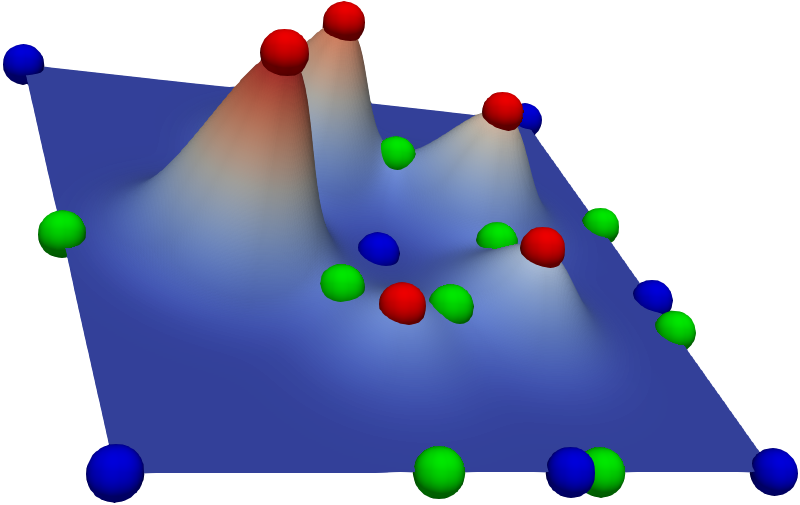}
\label{fig:crit}}
~
\subfigure[MS complex]{\includegraphics[width=0.21\textwidth]{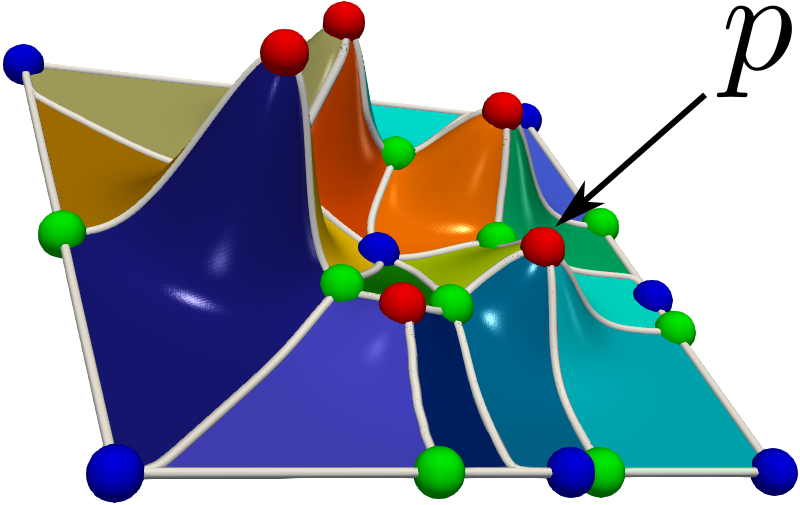}
\label{fig:ms1}}
~
\subfigure[MS complex, simplified]{\includegraphics[width=0.21\textwidth]{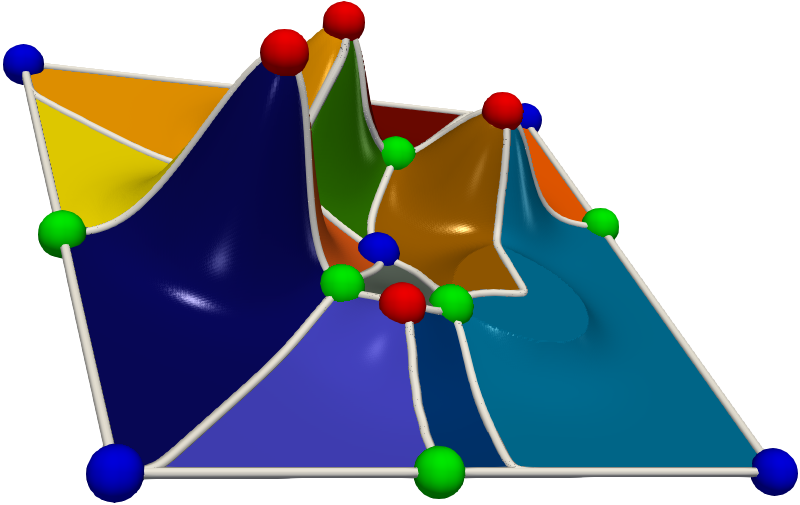}
\label{fig:ms2}}
~
\subfigure[Extremum graph]{\includegraphics[width=0.21\textwidth]{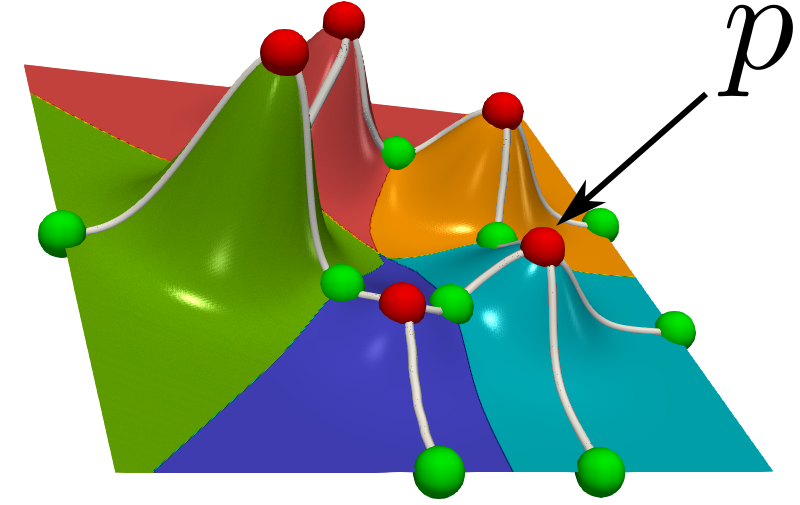}
\label{fig:ext1}}
~
\subfigure[Extremum graph, simplified]{\includegraphics[width=0.21\textwidth]{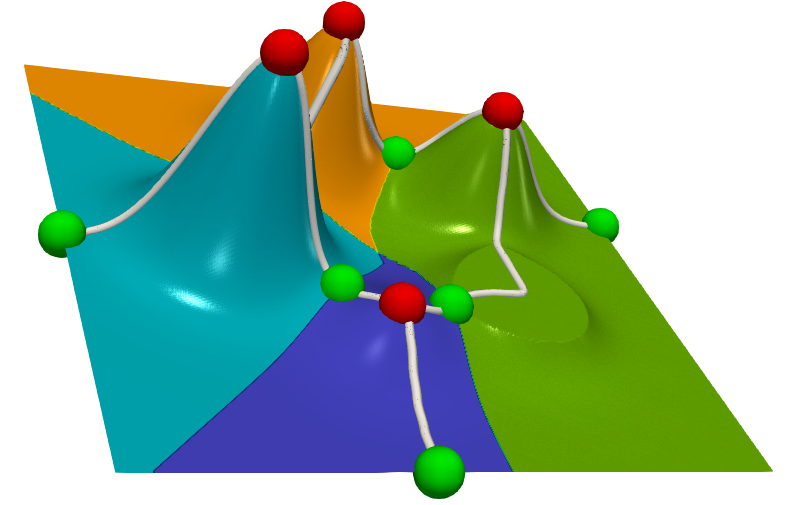}
\label{fig:ext2}}

\caption{\majorrevision{Extremum graph of a 2D scalar field. \subref{fig:field}~Scalar field $f$ defined on a 2D domain, shown using both a color map and surface height. \subref{fig:crit}~Critical points of $f$: maxima (red), saddles (green), and minima (blue). \subref{fig:ms1},\subref{fig:ms2}~MS complex and simplified MS complex obtained by canceling critical point pairs including $p$ and its adjacent saddle. The MS complex segments the domain into monotonic regions. \subref{fig:ext1},\subref{fig:ext2}~Extremum graph, original and simplified, of $f$ embedded within the domain represents the peaks and adjacency relationship between their corresponding segments. We consistently use the (\protect\includegraphics[height=0.15cm]{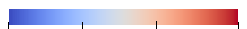}) color map for the scalar field.}}

\label{fig:illustration}
\vspace{-0.3in}
\end{figure}

Some of these topological structures have also been extended to time-varying fields~\cite{cohen2006vines,edelsbrunner2008time,oesterling2015computing}. Such a time-varying structure may be used for identifying and tracking the temporal evolution of the features and to analyze events such as creation and deletion, split and merge. Existing time-varying structures based on persistence diagrams and Reeb graphs need to be explicitly augmented in order to incorporate the geometric context. The MS complex includes such geometric context but they additionally represent a significantly large number of higher dimensional cells. Further, they are not stable in the presence of noise -- small variations in the input scalar field may result in significant changes in the complex because of their dependence on the gradient field. The extremum graph serves as a via media, because it captures the geometric context while the structure is small in size as it does not contain the higher dimensional cells unlike the MS complex. Figure~\ref{fig:illustration} shows a 2D scalar field whose topological features are represented by its critical points (maxima and minima). The extremum graph~\ref{fig:ext1} represents the peaks and the adjacency relationship between the regions associated with them. 

\begin{figure}

\centering
 
\subfigure[Time-varying field with extremum graphs]{\includegraphics[width=0.47\textwidth]{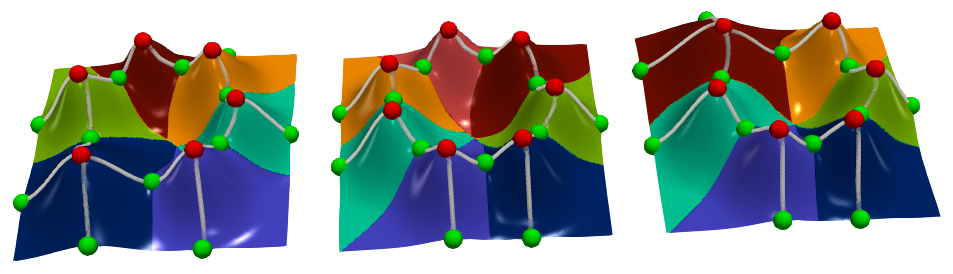}
\label{fig:eg}}
~
\subfigure[Temporal arcs of \TVEG]{\includegraphics[width=0.47\textwidth]{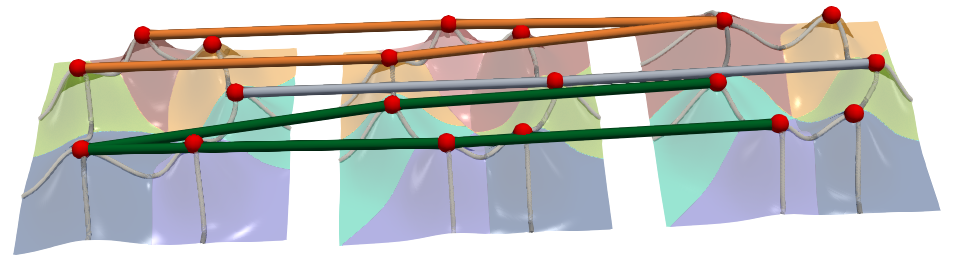}
\label{fig:tarc}}

\caption{\TVEG of a 2D time-varying scalar field. \subref{fig:eg}~Three time steps $f_1,f_2,f_3$ of a time-varying scalar field together with the respective extremum graphs $G^1,G^2,G^3$ embedded in the domain. Each peak is associated with a unique segment. Maxima are shown in red, saddles in green. \subref{fig:tarc}~A subset of the temporal arcs of the \TVEG. Green arcs correspond to a feature in $G^1$ that splits into two in $G^2$ and continues in $G^3$. The orange arcs correspond to two features in $G^1$ that continue onto $G^2$ but merge in $G^3$. The gray arcs correspond to a continuation of a feature from $G^1$ onto $G^3$. The remaining (gray) arcs that correspond to continuation of features and the saddle critical points are not shown for clarity. }

\label{fig:tvegillustration}
\vspace{-0.3in}
\end{figure}

In this paper, we present the time-varying extremum graph (\TVEG), an extension of extremum graph that facilitates visual analysis of time-varying scalar fields. \TVEG captures temporal events like creation/destruction, merge/split of topological features that are represented within the extremum graph at individual time steps. Figure~\ref{fig:tvegillustration} shows a 2D time-varying scalar field and \minoredits{its} associated extremum graph at each time step. The temporal arcs of the \TVEG correspond to split, merge, and continuation of a collection of topological features. \minorrevision{A monotonic path of temporal arcs is called a \TVEG \textit{track}. The extremum graphs endow the \TVEG tracks with a rich geometric context that supports the study of the track neighborhood and the analysis of topological features within the neighborhood.}

\subsection{Related work} \label{sec_retd_work}
\textbf{Extremum graphs.} The extremum graph was introduced by Correa et al~\cite{correa2011topological} as an intermediate step in the construction of topological spines, an augmented visual representations of topological features on the plane.  It plays a central role in the design of methods for comparative analysis, has been applied to detect periodicity and to track features in time-varying data~\cite{narayanan2015distance}, for symmetry detection~\cite{thomas2013}, and towards analysis and visualization of high dimensional functions~\cite{liu2019scalable}.

\noindent \textbf{Tracking based on topology of isosurfaces.} Many methods use isosurfaces as representatives for features of interest and track them. Shamir et~al.~\cite{shamir2002} describe a method to progressively track isosurfaces in time-varying scalar fields by combining spatial and temporal propagation followed by a step for tracking changes in topology. Sohn and Bajaj~\cite{sohn2005time} address the problem of finding correspondences between isosurfaces at a fixed isovalue across time. They do so by defining spatial overlaps between sublevel and superlevel sets and constructing a topology change graph (TCG). Both papers~\cite{shamir2002,sohn2005time} showcase the utility of the proposed method by tracking vortices in turbulent vortex data. More details regarding isosurface topology based tracking can be found in the comprehensive survey by Mascarenhas and Snoeyink~\cite{mascarenhas2009}. 

\noindent \textbf{Tracking based on topological structures.} Various methods have been proposed in the literature to track features captured by topological structures. Laney et al.~\cite{Laney2006} use the MS complex to define and represent bubbles in the mixing envelope of hydrodynamic instabilities and track them over time. Bremer et al.~\cite{Bremer2010} use the Morse complex to define features, followed by a hierarchical representation and construction of tracking graphs to track the evolution of combustion in lean premixed hydrogen flames. Bremer et al.~\cite{bremer2010interactive} facilitate exploration and analysis of burning cells from turbulent combustion simulation. Weber et al.~\cite{Weber2011} track burning regions by extracting isovolumes in a 4D space-time temperature field, followed by construction of Reeb graphs of time defined on the 4D domain. They convert a 4D tracking problem into the computation of the Reeb graph. Widanagamaachchi et al.~\cite{Widanagamaachchi2012} use correspondences between branches of merge trees followed by progressive construction of tracking graphs to track features in combustion simulations. %They also provide a linked view for concise and effective visualization of the features. 
In subsequent work~\cite{Widanagamaachchi2015}, they present methods to handle temporal artifacts, perform temporal simplification, and track embedded features via a parameter independent approach and apply it to track extinction holes in turbulent combustion simulations. Saikia and Weinkauf~\cite{saikia2017} use merge trees to represent multi-scale features and use a global shortest path formulation together with dynamic time warping to identify similar spatio-temporal structures. Lukasczyk et al.~\cite{lukasczyk2017,lukasczyk2019} use a nesting tree,  a variant of the merge tree, to capture hierarchical features and track them in dynamic nested tracking graphs. Soler et al.~\cite{soler2018lifted} use lifted Wasserstein matcher to find temporal correspondences between critical points and track features. Many other works propose different forms of tracking graphs to support the study of evolving topological features~\cite{kopp2018temporal,schnorr2018feature,schnorr2019feature,meduri2024jacobi}.
\minorrevision{Tracking features in vector fields is also an active area of research~\cite{tricoche2002topology,theisel2003feature,garth2004tracking,reininghaus2011efficient}}.

\noindent \textbf{Time-varying topological structures.} Many time-varying counterparts of topological structures have been proposed to facilitate analysis of feature rich data. Cohen-Steiner et al.~\cite{cohen2006vines} describe an algorithm to update persistence diagrams and use it to study protein folding trajectories.  Edelsbrunner et al.~\cite{edelsbrunner2008time} apply Jacobi curves~\cite{edelsbrunner2002jacobi} to track the temporal evolution of Reeb graphs by describing a complete characterization of the combinatorial changes that occur in the Reeb graph of a time-varying scalar field. Oesterling et al.~\cite{oesterling2015computing} introduced time-varying merge trees, which provides a topological summary of time-varying scalar fields, and showcase its utility with an application to analysis of time-varying high dimensional point clouds.  Wang et al.~\cite{wang2013visualizing} describe methods to robustly track critical points in 2D time-varying vector fields. They introduce two notions of robustness, static and dynamic, to understand temporal stability of critical points and provide tools to visualize them.

The above-mentioned methods and topological structures are based on isosurfaces, critical points, persistence diagrams, and variants of Reeb graphs (namely, Reeb graphs, merge trees, and contour trees). Following Yan et al.~\cite{yan2021survey}, they may be categorized as diagram based or graph based structures. 
\majorrevision{The diagram and graph based structures do not consider geometry, which is crucial in many applications, in contrast to complex based structures such as the MS complex or the extremum graph. The extremum graph captures the spatial connections between a maximum and all saddles in the neighborhood, has a natural embedding in the domain, and can be computed efficiently. Correa et al.~\cite{correa2011topological} demonstrate via examples, the additional geometric structure that is captured in the extremum graph as compared to the contour tree.}
\minorrevision{There are no time-varying extensions for complex based structures for scalar fields~\cite{yan2021survey}. Previous work on vector field topology have considered the time-varying scenario and tracked critical points along with saddle connections~\cite{theisel2004stream}. This paper attempts to take a step towards filling the gap in the context of scalar fields by introducing a time-varying structure based on extremum graphs.}

\subsection{Contributions}
We introduce the time-varying extremum graph (\TVEG), a topological structure for representing the combinatorial structure of the Morse decomposition of a scalar field and its evolution over time. We discuss its applications to feature exploration and tracking. While tracking is an important application, \TVEG is not limited to tracking. It is a data structure that stores a dynamic graph with potential for enabling various applications and supporting a wide variety of queries. Key contributions of this paper include
\begin{itemize}
\item A definition of a novel topological structure, the time-varying extremum graph (\TVEG).
\item An algorithm for constructing the \TVEG based on a formulation as an optimization problem.
\item Application to the study of interesting features and topological events in synthetic and simulation datasets that demonstrate the utility of the topological structure. 
\end{itemize}

%% file: section2.tex
\section{Background} \label{sec_background}
In this section, we introduce the relevant background on Morse functions and MS complex that will be required to define the time-varying extremum graph and to describe methods for computing the graph.

\subsection{Morse-Smale complex}
A Morse-Smale (MS) complex is a data structure that represents topological features in a scalar field by segmenting it into regions of uniform gradient flow, see  Figure~\ref{fig:ms1}. Formally, given a function $f:M\rightarrow{\mathbb{R}}$ defined on an $n$-dimensional manifold $M$ representing the scalar field, a \emph{critical point} of $f$ is a point where the gradient becomes zero, $\nabla f=0$. Other points of the domain are called \emph{regular}. Consider a maximal curve $\xi(r)$ parameterized by $r$ that satisfies $\frac{d}{dr}\xi(r)=\nabla f(\xi(r))$ at each point on $\xi$. Such a curve on $M$ passes through regular points and is called an \emph{integral line}. The limit points of an integral line are exactly the critical points of $f$. The collection of all integral lines that begin at a critical point $p$ is referred to as the \emph{ascending manifold} of $p$. Similarly, the collection of all integral lines that terminate at a critical point $p$ is referred to as the \emph{descending manifold} of $p$. A function $f$ is called a \emph{Morse function} if all its critical points are non-degenerate, \ie the Hessian matrix comprising the second order partial derivatives is nonsingular. The critical points of a Morse function $f$ are classified based on their \emph{Morse index}, which counts the number of negative eigenvalues of the Hessian (\ie the number of independent directions along which the function $f$ decreases). The index of a \emph{minimum} is $0$ and index of a \emph{maximum} is $n$. Other critical points are referred to as \emph{k-saddles}, where $1 \leq k \leq n-1$ is the Morse index. 

The ascending manifold of an index-$k$ critical point has dimension $n-k$ whereas its descending manifold has dimension $k$. The collection of ascending manifolds partition the domain of $f$ into valley-like regions. Similarly, the collection of descending manifolds partition the domain of $f$ into mountain-like regions. These partitions are called the \emph{Morse decomposition}. The overlay of the ascending and descending manifolds results in a partition of the domain of $f$ into a collection of monotonic regions. Each segment in the overlay is defined as a collection of integral lines that share a common origin and destination. For example, in Figure~\ref{fig:ms1}, the collection of integral lines that begin at a minimum and terminate at a maximum constitutes a 2-dimensional cell. The MS complex represents the combinatorial structure of the overlay. A $k$-dimensional cell of the complex represents a monotonic region of the overlay. The MS complex of a 2-dimensional Morse function consists of 0-, 1-, and 2-cells. We refer to the 0-cells as \emph{nodes} and 1-cells as \emph{arcs}.

The MS complex provides a comprehensive description of the topology of a scalar field and hence used for several applications. However, the complexes are large and often not amenable to direct visualization and interactive exploration, even for medium sized data. Noise in the data manifests as spurious saddle points and an exponential increase in number of cells of the complex, causing further impediment to the generation of meaningful visualizations. 

\subsection{Topological simplification}
Topological noise can be removed by simplifying the MS complex using a sequence of critical point pair cancellation operations~\cite{edelsbrunner2003hierarchical}. The cancellation operations are scheduled based on the notion of persistence~\cite{edelsbrunner2002persistence}. Topological persistence quantifies the relative importance of a pair of critical points that are connected by an arc in the MS complex in terms of the absolute difference between their function values. The MS complex is simplified by repeatedly canceling critical point pairs based on the increasing order of their persistence. A cancellation operation removes the two critical points, the arc connecting them, arcs incident on the two critical points, and reconnects the surviving critical points in the neighborhood~\cite{edelsbrunner2003hierarchical,gyulassy2006simplification,shivashankar2015felix,pandey2022morse}. The MS complex in Figure~\ref{fig:ms1} is simplified by canceling a maximum-saddle pair, resulting in the MS complex in Figure~\ref{fig:ms2}. The simplification removes the peak $p$ and merges its associated segment with the adjacent \minoredits{segments}.

\subsection{Extremum graph}
As observed by Correa et al.~\cite{correa2011topological}, interesting and meaningful topological structures within a scalar field are often associated with extrema. This observation serves as a motivation to introduce a topological abstraction called the extremum graph. The \emph{extremum graph} is a substructure of MS complex that captures the connectivity between maxima and saddles (maximum graph) or between minima and saddles (minimum graph). In this paper, we focus on the maximum graph while referring to it as the extremum graph to simplify terminology. The node set $V$ of an extremum graph $G(V,E)$ consists of the maxima and $n-1$-saddles of $f$. The set $E$ consists of arcs $(m_i,s_j)$ between maxima and $n-1$-saddles. If a pair of maxima $m_1,m_2 \in V$ are adjacent to a common saddle $s_1$, namely both $(m_1,s_1)$ and $(m_2,s_1)$ belong to $E$, then the segments associated to $m_1$ and $m_2$ are adjacent to each other. The extremum graph may be simplified by simplifying the corresponding MS complex. Figure~\ref{fig:ext1} shows the extremum graph of a 2D scalar field and Figure~\ref{fig:ext2} shows the simplified extremum graph obtained by canceling a maximum-saddle pair.

%% file: section3.tex
\section{\TVEG : definition and construction} \label{sec_tveg}
We describe a design of \TVEG for representing the temporal dynamics of a time-varying scalar field $\mathcal{F}$. The design aims to capture the temporal dynamics of both individual topological features and of a cluster of features that lie within a spatial neighborhood. We require that temporal slices of the \TVEG at all sampled time steps of $\mathcal{F}$ result in an extremum graph. The node set of the \TVEG consists of maxima and $(n-1)$-saddles of $\mathcal{F}$ at every time step. An arc in the \TVEG either belongs to an extremum graph of $\mathcal{F}$ restricted to a given time step or is a temporal arc that connects nodes between two consecutive time steps.

\begin{figure}
\centering
\includegraphics[width=0.37\textwidth]{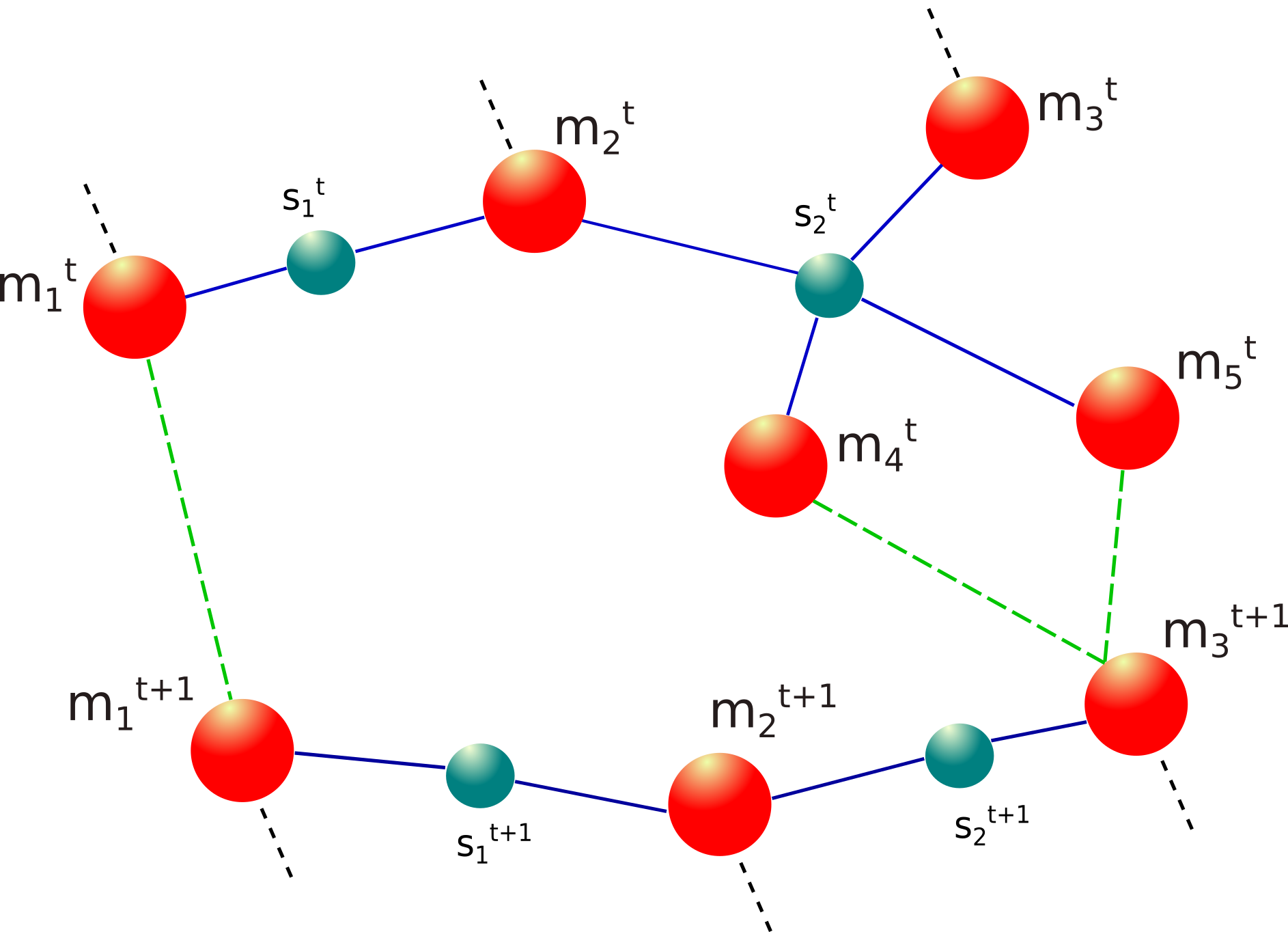}
\caption{\TVEG nodes and arcs. A subset of a \TVEG between two time steps $t$ and $t+1$. Sets $\{m_1^t,\ldots,m_5^t\}$ and $\{s_1^t,s_2^t\}$ consists of maxima and $(n-1)$-saddles from the extremum graph at time step $t$. Sets $\{m_1^{t+1},m_2^{t+1},m_3^{t+1}\}$ and $\{s_1^{t+1},s_2^{t+1}\}$ are maxima and saddles of the extremum graph at time step $t+1$. Arcs of the extremum graph are shown in blue, temporal arcs as green dashed edges, and connections to nodes outside the figure are shown as black dashed edges. Maxima $m_2^t$ and $m_3^t$ do not have a suitable temporal correspondence and hence die at time step $t$ whereas $m_2^{t+1}$ is born. Maxima $m_4^t$ and $m_5^t$ merge into $m_3^{t+1}$.}
\label{fig_toy_tveg}
\vspace{-0.3in}
\end{figure}

Let $\mathcal{G}^t(V^t,E^t)$ denote the extremum graph of $\mathcal{F}$ at a given time step $t$, $1 \leq t \leq T$. The \TVEG ${\mathcal{G}^*}(V^*,E^*)$ is a graph that contains all $\mathcal{G}^t$ as subgraphs in addition to arcs between nodes from consecutive time steps. 
Thus, the node set $V^*$ can be expressed as $V^* = \bigcup_{t=1}^T V^t = \{\bigcup_{t=1}^T M^t,\bigcup_{t=1}^T S^t\}$. We further distinguish between the maxima and $(n-1)$-saddles of $V^t$, and collect them into sets $M^t$ and $S^t$, respectively.  The arc set $E^*$ can be expressed as $E^* = \{ \bigcup_{t=1}^T E^t, \bigcup_{t=1}^{T-1}A^t \}$, consisting of two types of arcs namely the extremum graph arcs $E^t$ and the temporal arcs $A^t$. An arc in $E^t$ connects a maximum and a $(n-1)$-saddle. A temporal arc in $A^t$ represents the correspondence between two maxima that lie in extremum graphs of two consecutive time steps. In other words, an element $e = (m_i^t,m_j^{t+1}) \in A^t$ represents a correspondence between maxima $m_i^t \in M^t$ and $m_j^{t+1} \in M^{t+1}$. \majorrevision{As a design consideration, for each maximum $m_i^t \in M^t$ we allow at most two temporal correspondences with maxima in time step $t+1$ with the goal of allowing the representation of split events. The selection of the temporal arcs is formally described in Section~\ref{sec_tveg_temporal_arcs}.}

Figure~\ref{fig_toy_tveg} shows a subgraph of a \TVEG consisting of temporal arcs between time steps $t$ and $t+1$. Note that $A^t$ does not include temporal arcs between saddles. Maxima exhibit relatively higher stability over time in terms of spatial movement when compared to saddles. So, we chose to restrict temporal correspondence to those between maxima. The arcs set $A^t$ of $\mathcal{G}^*$ can be effectively used to represent the temporal variation in $\mathcal{F}$ together with topological events such as split, merge, generation, and deletion of features. \minorrevision{In a given time step $t$, critical points that have no correspondences with critical points in time steps $t-1$ and $t+1$ constitute \emph{generation} and \emph{deletion} events, respectively. A critical point from time step $t$ that has multiple correspondences with critical points in time step $t-1$ or $t+1$ constitutes a \emph{merge} or \emph{split} event, respectively.}

Figure~\ref{fig:tvegillustration}  illustrates how the \TVEG captures the different events in a 2D time-varying scalar field. Arcs in $E^t$ (light-gray) connect maxima with saddles within a time step. Arcs in $A^t$ (green, orange, gray) highlight temporal arcs that capture different topological events.

\subsection{Temporal correspondence}\label{sec_tveg_temporal_arcs}

A temporal arc represents the correspondence between a pair of maxima from two consecutive time steps. We now elaborate upon an optimization criterion that determines the temporal arcs. 

\majorrevision{We define the temporal arc set $A^t$ as the maximum cardinality set of arcs between maxima $m_i^t$ in time step $t$ and their correspondences in time step $t+1$ that satisfy some constraints, see Equation~\ref{eqn:definitin_A_t}. Equation~\ref{eqn:top-two-correspondences} describes correspondences between a maximum $m_i^t$ and two maxima, $m_{j_1}^{t+1}$ and $m_{j_2}^{t+1}$, from time step $t+1$. The scores associated with the two correspondences are expressed as the smallest and second smallest value of an objective function over all maxima $m_j^{t+1} \in M^{t+1}$ subject to structural constraints listed in Equation~\ref{eqn:z-merge-split-constraint}. \minorrevision{The objective function for the scores consists of four components -- functional persistence of the maxima $\mathcal{P}$, function value at maxima $\mathcal{J}$, spatial separation between the maxima $\mathcal{D}$, and neighborhood $\mathcal{N}$ of the maxima within the respective extremum graphs $\mathcal{G}^t$ and $\mathcal{G}^{t+1}$. The structural constraints disallow `z'-shaped configurations, thereby ensuring that no arc participates simultaneously in both a merge and split event.}
%
%%%%%%%%%%%%%%%%%%%%%%%%%%%%%%%%%%%%%
% \begin{equation}
% \begin{split}
% & \max |A^t|, \text{ where} \\
% A^t &\subseteq \bigcup_{m_i^t \in M^t} \{ (m_i^t,m_{j_1}^{t+1}), (m_i^t,m_{j_2}^{t+1}) \} 
% \end{split}
% \end{equation}
% \begin{equation}
% \begin{split}
% m_{j_1}^{t+1} = \argmin_{m_j^{t+1} \in M^{t+1}} \mathcal{P}(m_i^t,m_j^{t+1}) + \mathcal{J}(m_i^t,m_j^{t+1}) + \\ \mathcal{D}(m_i^t,m_j^{t+1}) + \mathcal{N}(m_i^t,m_j^{t+1})}\\~\\
% m_{j_2}^{t+1} = \argmin_{m_j^{t+1} \in M^{t+1}\setminus{m_{j_1}^{t+1}}} \mathcal{P}(m_i^t,m_j^{t+1}) + \mathcal{J}(m_i^t,m_j^{t+1}) + \\ \mathcal{D}(m_i^t,m_j^{t+1})  + \mathcal{N}(m_i^t,m_j^{t+1})
% \label{eqn:top-two-correspondences}
% \end{split}
% \end{equation}
% \begin{equation}
% \begin{split}
% & \text{subject to the constraint that} \\
% & (m_k^t,m_{j_1}^{t+1}) \not\in A^t \text{ and } (m_k^t,m_{j_2}^{t+1}) \not\in A^t \text{ for all }  m_k^t \in M^t\setminus{m_i^t}
% \label{eqn:z-merge-split-constraint}
% \end{split}
% \end{equation}
%
%%%%%%%%%%%%%%%%%%%%%%%
\begin{equation}
\begin{split}
\quad & \max |A^t|, \text{ where} \\
A^t &\subseteq \bigcup_{m_i^t \in M^t} \{ (m_i^t,m_{j_1}^{t+1}), (m_i^t,m_{j_2}^{t+1}) \} 
\label{eqn:definitin_A_t}
\end{split}
\end{equation}
\begin{equation}
\begin{split}
m_{j_1}^{t+1} = \quad \argmin_{m_j^{t+1} \in M^{t+1}} \quad (
 &\mathcal{P}(m_i^t,m_j^{t+1}) + \mathcal{J}(m_i^t,m_j^{t+1}) + \\ 
& \mathcal{D}(m_i^t,m_j^{t+1}) +  \mathcal{N}(m_i^t,m_j^{t+1})  ) \\
& \quad \\
m_{j_2}^{t+1} =  \argmin_{m_j^{t+1} \in M^{t+1}\setminus{m_{j_1}^{t+1}}} ( &\mathcal{P}(m_i^t,m_j^{t+1}) + \mathcal{J}(m_i^t,m_j^{t+1}) + \\ 
& \mathcal{D}(m_i^t,m_j^{t+1})  + \mathcal{N}(m_i^t,m_j^{t+1}) )
\label{eqn:top-two-correspondences}
\end{split}
\end{equation}
\begin{equation}
\begin{split}
& \text{subject to the constraint that} \\
& (m_k^t,m_{j_1}^{t+1}) \not\in A^t \text{ and } (m_k^t,m_{j_2}^{t+1}) \not\in A^t \text{ for all }  m_k^t \in M^t\setminus{m_i^t}
\label{eqn:z-merge-split-constraint}
\end{split}
\end{equation}

}
%%%%%%%%%%%%%%%%%%%%%%%
\majorrevision{As a maximum cardinality set, $A^t$ includes all temporal arcs defined in Equations~\ref{eqn:definitin_A_t} and~\ref{eqn:top-two-correspondences} that satisfy the constraint in Equation~\ref{eqn:z-merge-split-constraint}.} We allow for two correspondences between $m_i^t$ and maxima in time step $t+1$, thereby supporting the representation of split events. \majorrevision{Due to this design consideration, $m_i^t$ will be incident on at most two edges from $A^t$.}
The framework is designed to support larger number of correspondences per maximum if desired. However, a larger number of correspondences results in increased complexity of the resulting \TVEG tracks, \majorrevision{which may be difficult to analyze and cause visual clutter. The extremum graphs restricted to each time step are an important component of \TVEG. These extremum graphs} are simplified prior to \TVEG construction, which removes critical point pairs that are functionally similar and hence reduces the number of potential correspondences.
 
The persistence component $\mathcal{P}(m_i^t,m_j^{t+1})$ of the objective function is equal to the absolute difference between the topological persistence of $m_i^t$ and $m_j^{t+1}$. 
The functional component $\mathcal{J}(m_i^t,m_j^{t+1}) = \lvert f(m_i^t) - f(m_j^{t+1})\rvert$ measures the absolute difference between function values at the maxima. The distance component $\mathcal{D}(m_i^t,m_j^{t+1})$ is equal to the Euclidean distance between the maxima. The neighborhood component $\mathcal{N}(m_i^t,m_j^{t+1}) = \lvert \eta(m_i^t) - \eta(m_j^{t+1})\rvert$, where $\eta(m)$ for a maximum $m$ is defined as $\eta(m) = \sum_{i\in N(m)} \lvert f(m) - f(i)\rvert $ \ie the sum of absolute difference of function values between the maximum $m$ and all saddles in its neighborhood $N(m)$. While the first three components depend exclusively on the maxima and the topological feature that they represent, the component $\mathcal{N}$  captures the differences between the local connectivity of the maxima in consecutive steps. Hence, this component is crucially dependent on the extremum graph.
Topological persistence is a good stable measure of the topological feature represented by a maximum-saddle pair, and hence included in the objective function. The functional and spatial components are local properties of the maxima that  incorporates finer grained analysis of similarity and correspondence between consecutive time steps. Finally, the fourth component helps capture topological similarity in terms of the neighborhood of the maxima within the respective extremum graphs.

In order to reduce the effect of noise on the score, low persistent maxima are canceled in a first step based on a user specified persistence threshold and the objective function is computed post-simplification~\cite{edelsbrunner2002persistence,edelsbrunner2003hierarchical}. Critical point pair cancellations are performed independently within the two time steps. 

\begin{algorithm}
\SetAlgoLined
\DontPrintSemicolon
\SetKwInOut{Input}{Input}\SetKwInOut{Output}{Output}
\Input{A set of extremum graphs $[\mathcal{G}^p, \ldots, \mathcal{G}^r]$}
\Output{Temporal arc set $A^{t*}$ \\ Topological event sets $\mathcal{E}^{m*}, \mathcal{E}^{s*}, \mathcal{E}^{d*}, \text{and } \mathcal{E}^{g*}$}
\BlankLine
\textbf{Initialization:} $A^{t*} \leftarrow \varnothing$; $A^t \leftarrow \varnothing$; $M^0 \gets \varnothing$; $M^1 \gets \varnothing$; \; 

\tcc{Initialize $M^0$ as maxima set of $\mathcal{G}^p$}
$M^0 \gets M^p$

\tcc{Initialize all topological event sets}
$\{\mathcal{E}^{m*}, \mathcal{E}^{s*}, \mathcal{E}^{d*}, \mathcal{E}^{g*}\} \gets\varnothing$\;
\For{$i \gets p+1\; \KwTo \; r$}{

    \tcc{Initialize $M^1$ as maxima set of $\mathcal{G}^i$}
    $M^1 \gets M^i$

    $\mathcal{S} \gets \textsc{ComputeScores}(M^0,M^1)$\;
    $\mathcal{S} \gets \textsc{FilterScores}(\mathcal{S})$\;
    \tcc{Compute the temporal arc set $A^t$}
    \ForEach{$(m^0,m^1,s) \in \mathcal{S}$}{
        $A^t\gets A^t \cup (m^0,m^1)$\;
    }
    \tcc{Detect topological events}
    $\mathcal{E}^m \gets \textsc{DetectMerge}(\mathcal{S},i)$\; 
    $\mathcal{E}^s \gets \textsc{DetectSplit}(\mathcal{S},i)$\;
    $\mathcal{E}^d \gets \textsc{DetectDel}(\mathcal{S},M^0,i)$\; 
    $\mathcal{E}^g \gets \textsc{DetectGen}(\mathcal{S},M^1,i)$\;
    \tcc{Remove z-shape configurations.}
    $\mathcal{W}\gets\mathcal{E}^m \bigcap \mathcal{E}^s$\;
    \Repeat{$\mathcal{W}=\emptyset$}
    {$w \gets \textsc{MaxScoreEdge}(\mathcal{W}$)\;
    $A^t\gets A^t \setminus w$\;
    $\mathcal{E}^m,\mathcal{E}^s  \gets\textsc{UpdateMergeSplitEdges}(\mathcal{E}^m,\mathcal{E}^s,w$)\;
    $\mathcal{W}\gets\mathcal{E}^m \bigcap \mathcal{E}^s$\;}
    \tcc{Populate temporal arc set}
    $A^{t*}\gets A^{t*}\cup A^t$\;
%    \tcc{Update conflicts in merge/split sets}
%    $\mathcal{E}^m\gets \mathcal{E}^m \setminus \mathcal{W}$, 
%    $\mathcal{E}^s\gets \mathcal{E}^s \setminus \mathcal{W}$\;
    \tcc{Update topological event sets}
    $\mathcal{E}^{m*}\gets\mathcal{E}^{m*}\cup\mathcal{E}^m$\;
    $\mathcal{E}^{s*}\gets\mathcal{E}^{s*}\cup\mathcal{E}^s$\;
    $\mathcal{E}^{d*}\gets\mathcal{E}^{d*}\cup\mathcal{E}^d$\; 
    $\mathcal{E}^{g*}\gets\mathcal{E}^{g*}\cup\mathcal{E}^g$\;
    \tcc{Re-initialize for next iteration}
    $A^t\gets\varnothing$, $M^0\gets M^1$\;
}
\caption{\textsc{TemporalArcs}}
\label{algo_tem_arcs}
\end{algorithm}

\begin{table}
    \centering
    \caption{List of different attributes of the critical points.}
    \label{table_cp_fields}
    \begin{tabular}{r|l}
         \textbf{Fields} & \textbf{Description} \\
         \hline 
         \textit{id} & A unique id assigned to the critical point \\
         \textit{index} & Index\\ 
         $\Bar{x}$ & Coordinates\\ 
         \textit{pers} & Topological persistence \\
         $\eta$ & Neighborhood contribution\\
         \textit{ascmfold} & Ascending manifold\\
         \textit{dscmfold} & Descending manifold \\
         \textit{geom} & Ascending / Descending manifold geometry \\
         $t$ & The time stamp of the scalar field
    \end{tabular}
\end{table}
\subsection{Computation}\label{sec_tveg_computation}
We compute \TVEG in two steps. The first step constructs the extremum graphs for all the time steps of the time-varying scalar field $\mathcal{F}$ by either employing the approach of Correa et al.~\cite{correa2011topological} or computing the MS complex~\cite{shivshankar2012ms3dparallel,delgado2014skeletonization,bhatia2018topoms} and extracting the substructure. The topology toolkit (TTK)~\cite{tierny2017topology} also provides an implementation of the algorithm to compute MS complex and its substructures based on a discrete Morse theoretic approach~\cite{shivshankar2012ms3dparallel}. In either case, the graph is simplified using a persistence threshold \minoredits{$\theta$} to remove noise. Corresponding to each node of the extremum graph, we store a tuple that contains attributes related to the critical point (see Table~\ref{table_cp_fields}). An arc connecting a pair of critical points $v^1,v^2$ is stored in $E^t$ as $(v^1.id, v^2.id)$. 

The second step computes the correspondences between maxima from consecutive time steps. We now describe the algorithm \textsc{TemporalArcs} (Algorithm~\ref{algo_tem_arcs}) that computes and records temporal arcs of the \TVEG for all pairs of consecutive time steps over a given range $1 \le p < r \le T$. Apart from computing temporal arcs, it also detects and records topological events \minorrevision{across all time steps} such as merge, split, deletion, and generation in the sets $\mathcal{E}^{m*}$, $\mathcal{E}^{s*}$, $\mathcal{E}^{d*}$ and $\mathcal{E}^{g*}$, respectively. \textsc{TemporalArcs} begins by initializing the set $A^{t*}$ that stores all temporal arcs, sets $M^0$ and $M^1$ that store the collection of maxima in the two time steps, and sets for recording topological events. Iterating over pairs of consecutive time steps (lines 4-\minoredits{28}), the algorithm computes the correspondence scores for all maxima using \textsc{ComputeScores} (Algorithm~2). The scores are recorded in $\mathcal{S}$, and further refined using \textsc{FilterScores} (Algorithm~3) based on a threshold derived from the variance of the scores. The score computation and refinement is described later in this section. After refinement, $\mathcal{S}$ records the set of correspondences and scores and set $A^t$ is updated with the temporal arcs. 
 
\minorrevision{Next, all topological events between time steps $t$ and $t+1$ are recorded in sets $\mathcal{E}^m, \mathcal{E}^s, \mathcal{E}^d$, and $\mathcal{E}^g$ (lines 11-14). $\mathcal{E}^m, \mathcal{E}^s, \mathcal{E}^d$, and $\mathcal{E}^g$ correspond to merge, split, deletion and generation sets respectively. Algorithms 5-8 in the supplementary material compute these sets.} We then resolve any co-occurrences of merge and split or `z'-shaped configurations between two time steps (lines 15-21). These co-occurrences are recognized as a subset of the intersection of $\mathcal{E}^m$ and $\mathcal{E}^s$. This subset is computed using a greedy approach that iteratively removes the largest score edge in a 'z'-configuration from $A^t$. Finally, the set $A^{t*}$ containing  temporal correspondences over all iterations, the sets containing topological events per iteration, and the global containers for arcs and events are updated (lines 22-26). We refer the reader to the supplementary material for the descriptions of the individual subroutines.
 \textsc{TemporalArcs} computes the global temporal arc set of a \TVEG in quadratic time $\mathcal{O}(n^2)$, where $n = \max {|M^t|, p\leq t\leq r}$, the largest cardinality of the maxima sets over all time steps in the given input time range $[p,r]$.

\noindent \textbf{Score computation.}
The subroutine \textsc{ComputeScores} (Algorithm~2) uses Equation~\ref{eqn:top-two-correspondences} to compute correspondence scores between two sets of maxima $M^0$ and $M^1$. For each maximum $m^0 \in M^0$, it iterates exhaustively over all maxima in $M^1$, computes the objective function from Equation~\ref{eqn:top-two-correspondences} using the attributes (\emph{pers}, $\Bar{x}$, $\mathcal{F}(\Bar{x})$, $\eta$), and records them. The distance component is computed as the Euclidean distance between pair of maxima. The neighborhood component is computed by considering the local neighborhood contribution $\eta$ of the maxima. The two lowest scores are identified for $m^0$ and finally the output $\mathcal{S}$ containing the correspondences and their scores is returned as a set of tuples  $(m^0,m^1,s)$ of length three. In practice, to ensure that the addition of the four components in Equation~\ref{eqn:top-two-correspondences} is meaningful, we normalize them so that the values lie within $[0,1]$.

\noindent \textbf{Score refinement.}
The set of scores returned by \textsc{ComputeScores} (Algorithm~2) need to be further refined to assess whether a particular correspondence is meaningful and should be retained within $\mathcal{S}$. The refinement may lead to topological events such as deletion and generation. \textsc{FilterScores} (Algorithm~3) processes the input set $\mathcal{S}$ and refines the set based on a threshold. The procedure records all scores in a list $\mathcal{Y}$, computes the mean($\mu$) and the standard deviation($\sigma$) of scores, and sets the threshold $\tau$ to $\mu + \sigma$. Finally, all tuples corresponding to scores greater than or equal to $\tau$ are removed from $\mathcal{S}$. 

\noindent \minorrevision{\textbf{Optimal temporal arcs.}}
\minorrevision{The \TVEG computation is not set up as a global optimization problem. Instead, \TVEG computation progresses by focusing on a pair of consecutive time steps within a single iteration (Lines 4--27 of \textsc{TemporalArcs}). This focus implies that the identification of temporal arcs within a single iteration is not affected by and does not affect arcs computed between other pairs of time steps. Furthermore, the algorithm can be time efficient and scalable for datasets of larger sizes and complexity. Algorithm \textsc{TemporalArcs} is an exact implementation of the optimization criteria from 
 Equations~\ref{eqn:definitin_A_t}-~\ref{eqn:z-merge-split-constraint} and computes an optimal set of arcs.}

\subsection{\majorrevision{\TVEG supported queries}}
\majorrevision{The \TVEG serves as a data structure that can support various visualization and data related queries. These queries can be broadly classified as temporal or spatiotemporal in nature. For instance, a typical temporal query can help identify \TVEG tracks that are longer than a certain specified threshold. Topological events, specifically splits, merges, deletions, and generations can be quickly identified from simple temporal queries on \TVEG tracks. One spatiotemporal query helps identify the \TVEG tracks that have the least spatial deviation along time. These simple queries can help identify maxima that exhibit stable behavior along the \TVEG tracks and provide meaningful insights about their role in the temporal behavior of data. Additional queries include selection of tracks and the associated extremum graph neighborhood lying within a user specified spatial neighborhood or time interval.  

Topological events and how they affect the combinatorial structure of the extremum graph can help explain global changes within the data. 
Using spatiotemporal queries, one can observe structural changes within the extremum graph as it evolves over time. 
Further, identifying similar \TVEG tracks based on topological or geometric criteria followed by a clustering procedure helps study data dynamics. Observing groups of similar \TVEG tracks together with the extremum graphs helps identify and understand the correlation, if any, between the temporal changes in structure within extremum graph neighborhoods  and the global variations in data dynamics. This may be followed by a focused study of the temporal behavior of a particular neighborhood within the extremum graphs. Complex queries can be formed as a combination of the queries mentioned above, and hence support fine grained analysis of the data.}

\subsection{\minorrevision{\TVEG visualization}}

\minorrevision{The visualization of the \TVEG tracks consists of an overview via a representation of space-time and, optionally, direct embedding of extremum graphs in space for individual time steps. 
The overview is provided by a collective representation of all temporal arcs. In this representation, the 3D domain of the scalar field is scaled along the z-axis, and all instances of the domain over time are stacked along the z-axis to represent the space-time domain. Figure~\ref{fig_gaussian_tracks} shows one such scaled domain in the brown box together with the extremum graph at that time step. Additional examples can be found in Figures~\ref{fig_vf_tracks} and~\ref{fig_scw_comparison}. An extremum graph is considered as a basic unit in the visual representation of the \TVEG. %In addition to the vertices and edges in an extremum graph, \TVEG also includes temporal arcs that represent temporal correspondences. We adopt a simple strategy to organize the extremum graphs at every time step by stacking them along the z-axis of the data domain. Further, the temporal arcs are also oriented along the z-axis. 
As a consequence of this simple representation, the time axis coincides with the z-axis and the evolution of the extremum graph within the \TVEG is depicted along the z-axis. This design strategy helps reduce clutter in the visualization, which may occur if the temporal arcs do not align with one of the spatial axis. Scaling down the 3D domain along z-axis reduces the size of the spatial edges thereby rendering the temporal arcs to be visually prominent.}

%% file: section4.tex
\section{Case studies}\label{sec_case_studies}
We demonstrate the utility of \TVEG on a synthetic sum of 3D Gaussians dataset consisting of spatiotemporal movement of the centres of eight 3D Gaussians, a viscous fingers dataset~\cite{viscousfingers2016_dataset} that simulates the temporal mixing of salt and water resulting in finger like spatial structures, and a 3D B\'enard-von K\'arm\'an vortex street dataset~\cite{saikia2017,vonfunck08a}. The above-mentioned datasets are 3D time-varying scalar fields, so we use \textsc{pyms3d}~\cite{shivshankar2012ms3dparallel} to compute the extremum graph as a substructure of the MS complex for all time steps (see supplementary material). It is likely that multiple saddles are adjacent to a given pair of maxima. In all experiments, we retain the saddle with the highest function value to record the adjacency between the segments associated with the maxima, and discard the other saddles to reduce clutter. \minorrevision{The case studies demonstrate how the rich geometric context provided by the extremum graphs supports an effective analysis of the \TVEG tracks.}

The \TVEG is computed using the algorithm \majorrevision{\textsc{TemporalArcs}} described in Section~\ref{sec_tveg_computation}. \majorrevision{The time required for \TVEG computation would naturally depend on the complexity presented by each dataset. The average time taken to compute \TVEG for the 3D Gaussian, viscous fingers, and vortex street data are $0.92$~ms, $24.62$~ms and $18.67$~ms, respectively. A detailed report on the running times is available in the supplementary material. The extremum graph is assumed to be available as input for the \TVEG computation. \textsc{pyms3d} takes an average of $1.85$~s and $1.02$~s to compute the extremum graph for the 3D Gaussians and viscous fingers datasets, and takes $6.58$~s for the more complex vortex street dataset. Alternative methods are available to compute the extremum graph~\cite{ande2023tachyon} efficiently without explicit construction of the MS complex.}

\subsection{Moving Gaussians}\label{sec_gaussians} 
\begin{figure*}

 \centering
 
 \includegraphics[width=1\textwidth]{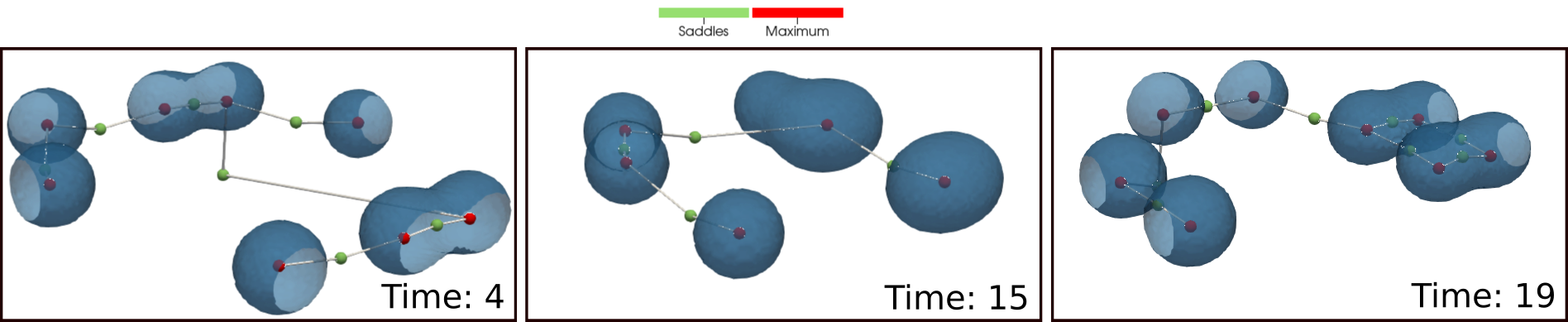}

  \caption{Dynamics in a synthetic sum of Gaussians dataset \textit{Gauss8}. The data is visualized by displaying the intersection of descending 3-manifolds of maxima with volume enclosed by the isosurface at scalar value 21. The resulting blobs merge over time to form larger components and subsequently split into multiple components. The merge and split behavior is also observable from the extremum graphs.} 

  \label{fig_gaussian}

\end{figure*}

The first case study consists of experiments on a synthetic dataset called \textit{Gauss8} defined on a $128\times 128\times 128$ grid over $50$ time steps. It represents the movement of eight 3D Gaussians whose centres move along predetermined paths while being restricted to the $x=0$ plane. The dataset is defined as the sum of eight 3D Gaussians, whose maxima are located exactly at the Gaussian centres. The trajectories of the centres are designed to induce  multiple topological events such as merges and splits. Figure~\ref{fig_gaussian} shows three time steps that exhibit splits and merges together with a change in the numbers and locations of maxima. 

\begin{figure*}

 \centering
 
 \includegraphics[width=1\textwidth]{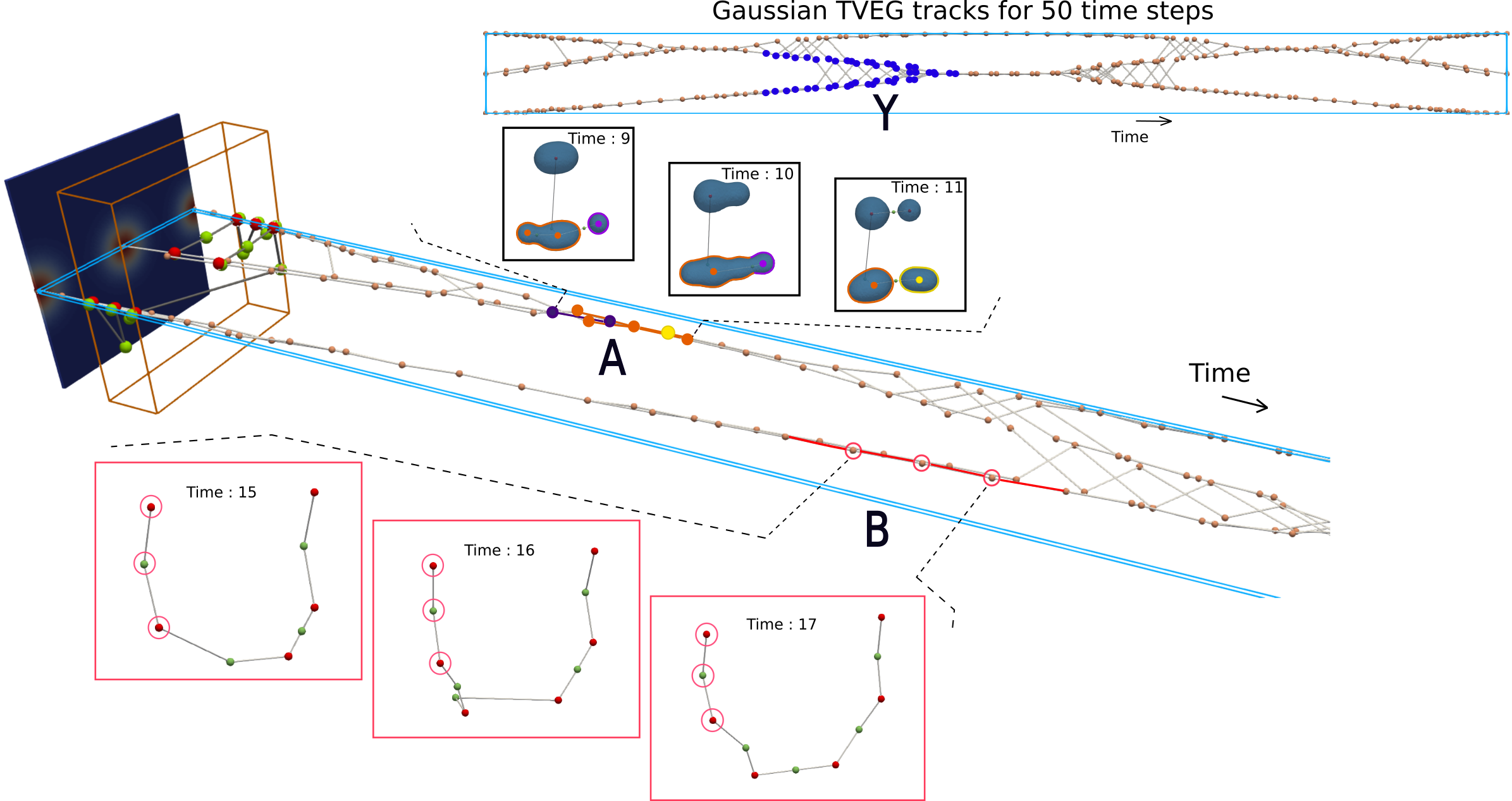}

  \caption{Temporal tracks from the \TVEG of \textit{Gauss8}. \majorrevision{The 3D domain is scaled along the z-axis and stacked to visualize the tracks over time. (top-right)~A view of the stacked domains from the top, along the y-axis, shows the tracks over all time steps.} A subset of \TVEG tracks (\textbf{Y}) that exhibits symmetry along time is highlighted in blue. (middle)~Track \textbf{A} is a subset of a longer track, consisting of time steps 9-11 and includes merge and split events. Inset depicts the blobs in the corresponding time steps 9, 10, and 11. Track \textbf{B} (red), a subset of \textbf{Y}, is selected to showcase the structural similarity between extremum graphs (inset) sampled at time steps 15, 16, and 17.} 

  \label{fig_gaussian_tracks}
\vspace{-0.3in}
\end{figure*}

Figure~\ref{fig_gaussian_tracks} shows an overview of \TVEG tracks. All maxima (red) in the extremum graph lie on the plane $x=0$ as expected and continue to be restricted to the plane over time, as is evident from the flat appearance of the temporal \TVEG tracks outlined by the cyan box. \majorrevision{We demonstrate the utility of \TVEG tracks via two tasks on \textit{Gauss8}.}

\begin{figure}

\centering
 
\subfigure[\TVEG tracks]{\includegraphics[width=0.21\textwidth]{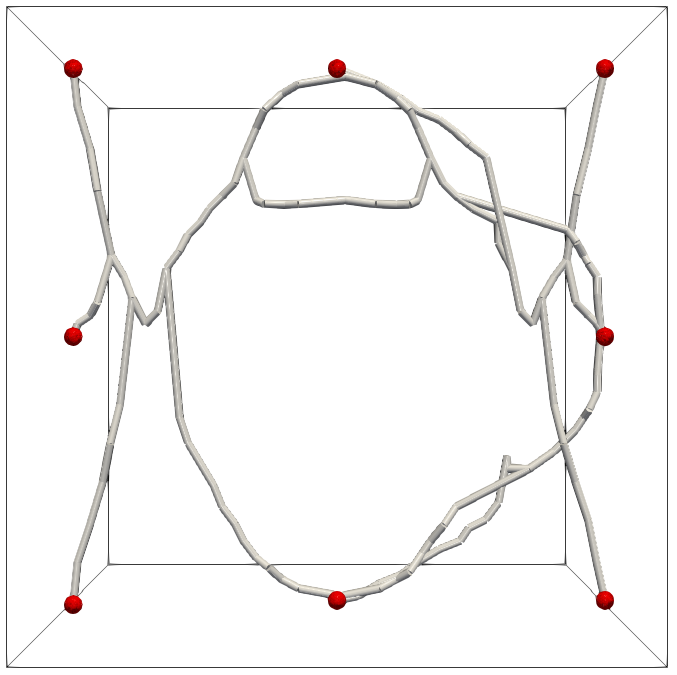}
\label{fig:comp1}}
~
\subfigure[\LWM tracks]{\includegraphics[width=0.21\textwidth]{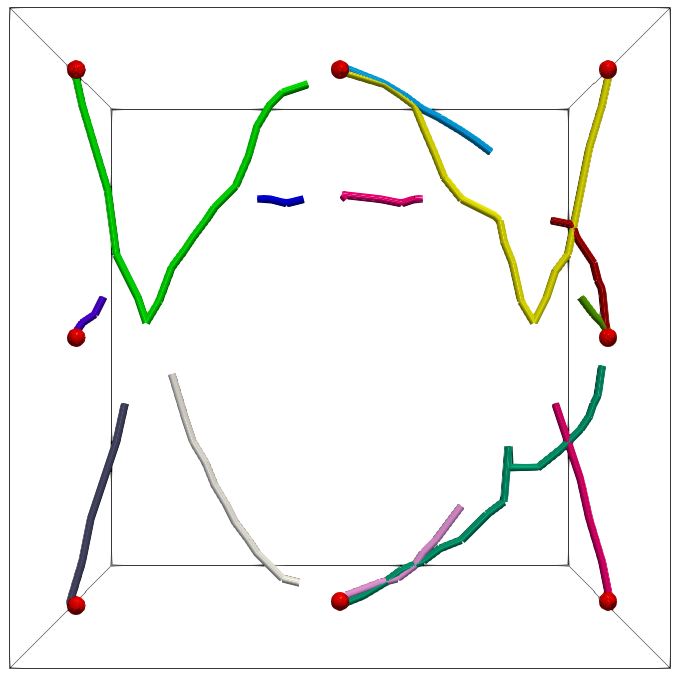}
\label{fig:comp2}}
~
\subfigure[\LWM track post processing, threshold = $8\%$]{\includegraphics[width=0.21\textwidth]{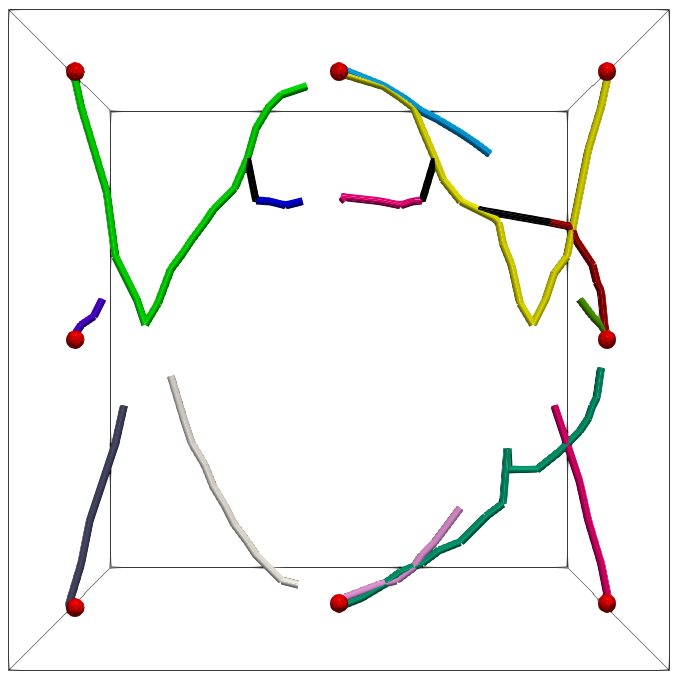}
\label{fig:comp3}}
~
\subfigure[\LWM tracks with post processing, threshold = $15\%$]{\includegraphics[width=0.21\textwidth]{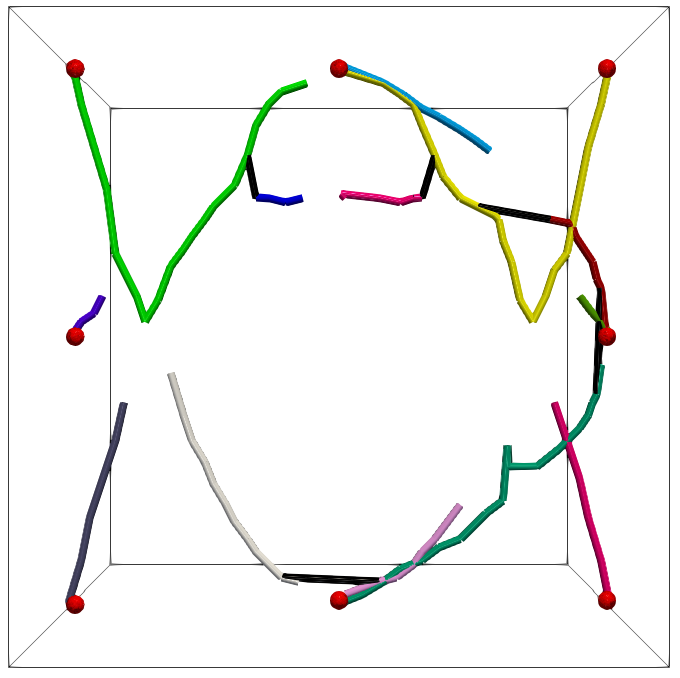}
\label{fig:comp4}}

\caption{\majorrevision{Tracks obtained by (a)~\TVEG and (b-d)~\LWM for the \textit{Gauss8} dataset. All maxima in the first time step are shown as red spheres. The postprocessing step is applied with two different distance thresholds, $0.08 \cdot diag$ and $0.15 \cdot diag$. Here, $diag$ is the length of the long diagonal of the data domain. Individual \LWM track components are shown in distinct color and the arcs inserted by the postprocessing step are shown in black.}}

\label{fig_lwm_comp}
\vspace{-0.3in}
\end{figure}
\noindent\textbf{Topological event detection.}
The merges and splits among the Gaussians can be visually identified from the \TVEG tracks in Figure~\ref{fig_gaussian_tracks}. From the overview of the temporal tracks (Figure~\ref{fig_gaussian_tracks}, top-right) computed over the entire time range of 50 steps, we  observe multiple merges followed by splits. We note an increasing number of cross correspondences as the tracks approach a merge event or subsequent to a split event. The existence of such cross temporal arcs between corresponding maxima over a time range can indicate gradually decreasing spatial distance between global components such as super-level sets eventually leading to topological merges/splits. A specific track segment of interest, \textbf{A}, is highlighted. The maxima merge, die, and are born at time steps 9,10,11. We choose an isosurface extracted at scalar value 21 to visualize the field. Each component of the isosurface associated with a maximum is identified by computing the intersection with the descending 3-manifold of the maximum. The blobs merge and subsequently split corresponding to the topological events in the extremum graph. Along track \textbf{A}, the maxima are colored to indicate their temporal correspondence. The corresponding maxima in the scalar domain and their associated isosurface components are also highlighted with the same color.  The purple maximum continues between time steps 9 and 10, whereas the orange maxima in time step 9 merge in time step 10. As a result of this interaction one component of the isosurface in time step 10 is jointly represented by the purple and orange maxima. The purple maximum vanishes between time steps 10 and 11 and a new yellow maximum appears, resulting in an isosurface component splitting into two. The \TVEG tracks help interpret the topological events.

\noindent\textbf{Similarity pattern within extremum graphs.}
The tracks of the Gaussian centres in \textit{Gauss8} are designed to be symmetrical as can be seen in the overview of the tracks over 50 time steps (Figure~\ref{fig_gaussian_tracks}, top-right). One such temporally symmetric region is labelled as \textbf{Y} (blue) within the overview. We studied structural changes within the extremum graphs from the time range when the \TVEG tracks exhibit symmetry. One such segment, track \textbf{B} (red), is highlighted to demonstrate the result of the study and three time steps (15, 16, and 17) from this segment are shown. Maxima that belong to track \textbf{B} are highlighted within the extremum graph using a red enclosing circle. One neighboring saddle and a maximum in the neighborhood of the saddle are also highlighted using a red circle. We observe a structural similarity between the highlighted regions of the extremum graphs in these time steps. This observation suggests that temporal geometric patterns within \TVEG tracks, if any, are indicative of a repetitive local structural pattern within the corresponding extremum graphs over time.

\majorrevision{\noindent\textbf{Comparison with Lifted Wasserstein Matcher.} The Lifted Wasserstein Matcher~(\LWM) provides an efficient method to track topological features via critical points by computing an optimal matching between critical points of the scalar field in two consecutive time steps~\cite{soler2018lifted}. The matching follows from a generalization of the Wasserstein distance between the persistence diagrams representing the two scalar fields.
The methodology for tracking employed in the \LWM is also based on a combination of topological and geometrical criteria. The \LWM is a good exemplar method for comparing results because of the similarity between the two approaches.
In addition to topological persistence and spatial location of the critical points, \TVEG incorporates local neighborhood information of the critical points and function value differences. 

The \LWM is implemented as a part of TTK~\cite{tierny2017topology} and is available as an open source software. We use TTK~0.9.9 to compute the tracks.
\LWM is executed using the recommended parameter values: extrema weight = 1.0, saddle weight 0.1, $x,y,z$ weights = 1.0,1.0,1.0. To facilitate comparison, we embed the tracks in the domain. To ensure a meaningful comparison, we also implement the postprocessing strategy discussed in the \LWM description but unavailable in the TTK implementation. Further, the \TVEG tracks are computed using only the top temporal correspondence, which does not capture the splits. 

We study the similarities and differences between the \TVEG tracks with those reported by \LWM, see Figure~\ref{fig_lwm_comp}. Overall, the set of tracks reported by the two methods are similar, which serves as a validation. The video in the supplementary material shows the \TVEG tracks annotated with the extremum graph for each time step. The isosurface at a fixed value of 20 (mid-point of the scalar range) is rendered to visually present the topological events.

The \TVEG tracks handle merge events directly. The latter half of the \textit{Gauss8} dataset is a time reversal of the first half. So, the split events in the first half become merge events in the latter half. The resulting collection of tracks form a single connected component. The set of \LWM tracks are disconnected prior to the postprocessing step because only birth/death events are detected. We implement the postprocessing that attempts to record merge/splits using a distance threshold that is defined as a fraction of the length of the long diagonal, $diag$, of the data volume. In Figures~\ref{fig:comp3} and~\ref{fig:comp4}, we show the results for two distance thresholds,  $8\%$ and $15\%$. We observe that some of the merge/split events (for example, as shown in the video in time step 20) are detected only for high values of the threshold and some breaks (for example, in time steps 3 and 8) continue to be present even when the threshold is high. This is perhaps because the critical points move significantly between time steps. The additional components in \TVEG, namely local neighborhood and function value, likely contribute to additional correspondences.}

\subsection{Viscous fingers}\label{sec_cs_viscous}
\begin{figure*}[!ht]

 \centering
 
 \subfigure[Arcs with Volume data]{\includegraphics[width=0.3\textwidth]{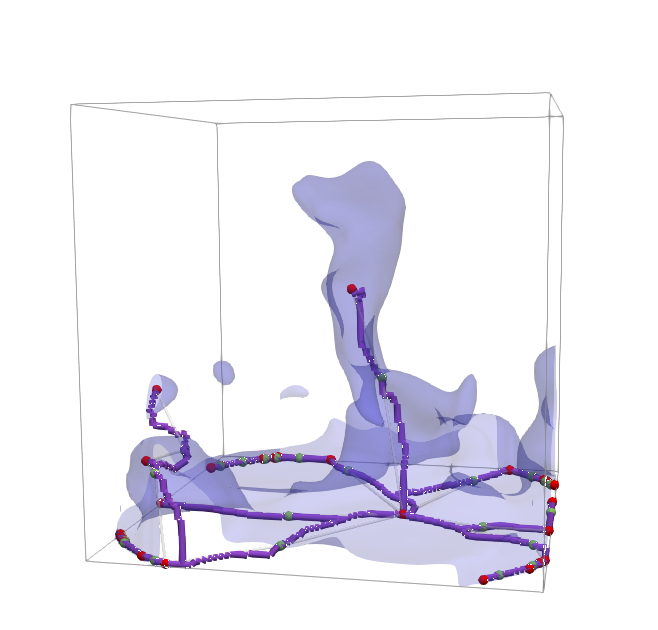}
\label{fig_ext_ms_data}}
~
\subfigure[Viscous Finger with 3D extremum graph]{\includegraphics[width=0.3\textwidth]{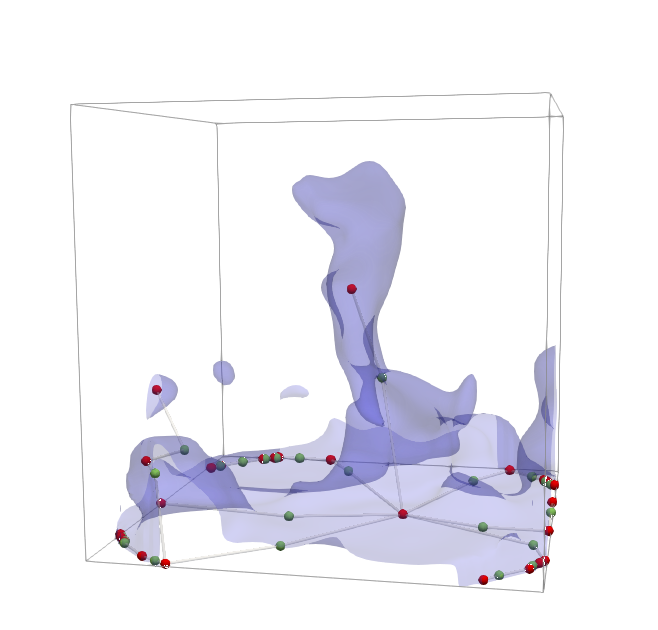}
\label{fig_ext_graph_data}}
~
\subfigure[3D Extremum graph with critical points]{\includegraphics[width=0.3\textwidth]{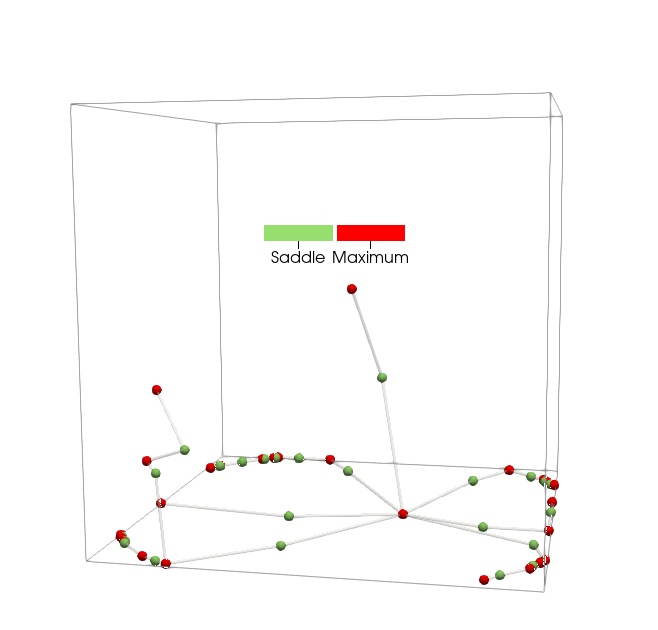}
\label{fig_ext_3dgraph}}
  \caption{Visualizing the viscous fingers dataset using the extremum graph. \subref{fig_ext_ms_data}~Fingers are visualized as intersection of descending 3-manifolds of salt concentration maxima and volume enclosed by isosurface at scalar value 35. Integral lines between each max-saddle arc in the extremum graph represent the skeletal structure of the fingers. \subref{fig_ext_graph_data}~All max-saddle arcs in the extremum graph are displayed using straight edges. \subref{fig_ext_3dgraph}~The extremum graph with maxima (red) and saddles (green).}
  \label{fig_eg_3d_graph}
  \vspace{-0.15in}
\end{figure*}
%
% \begin{figure*}

%  \centering
 
%  \subfigure{\includegraphics[width=0.18\textwidth]{fig/vf3d/34.png}
% \label{fig_vf34}}
% ~
% \subfigure{\includegraphics[width=0.18\textwidth]{fig/vf3d/49.png}
% \label{fig_vf49}}
% ~
% \subfigure{\includegraphics[width=0.18\textwidth]{fig/vf3d/70.png}
% \label{fig_vf70}}
% ~
% \subfigure{\includegraphics[width=0.18\textwidth]{fig/vf3d/72.png}
% \label{fig_vf72}}
% ~
% \subfigure{\includegraphics[width=0.18\textwidth]{fig/vf3d/98.png}
% \label{fig_vf98}}

%   \caption{Dynamics of Viscous finger formation is shown using the 3D data sampled at five time steps. Fingers are shown to grow gradually outward from the mixing interface between~\ref{fig_vf34} and~\ref{fig_vf70} and eventually leading to a split event~\ref{fig_vf72}}. 

%   \label{fig_vf_dynamics}

% \end{figure*}

Viscous fingers refer to a phenomenon that occurs during mixing of two fluids of different viscosity. We use a particle ensemble dataset~\cite{viscousfingers2016_dataset} generated using a stochastic simulation of the mixing process of high concentration salt into pure water. During the mixing process, high concentration finger like regions are formed within the medium that gradually grow away from the mixing interface and eventually terminate. The salt concentration data used in the following computational experiments is from the 33$^{rd}$ ensemble run at a smoothing length 20, and resampled from the original cylindrical domain onto a $101\times101\times101$ grid over $120$ time steps~\cite{favelier2016visualizing}. Figure~\ref{fig_eg_3d_graph} shows the extremum graph computed for this time-varying salt concentration scalar field at one time step. 

\begin{figure*}

 \centering
 
 \subfigure[Distance plot from mixing interface]{\includegraphics[width=0.9\textwidth]{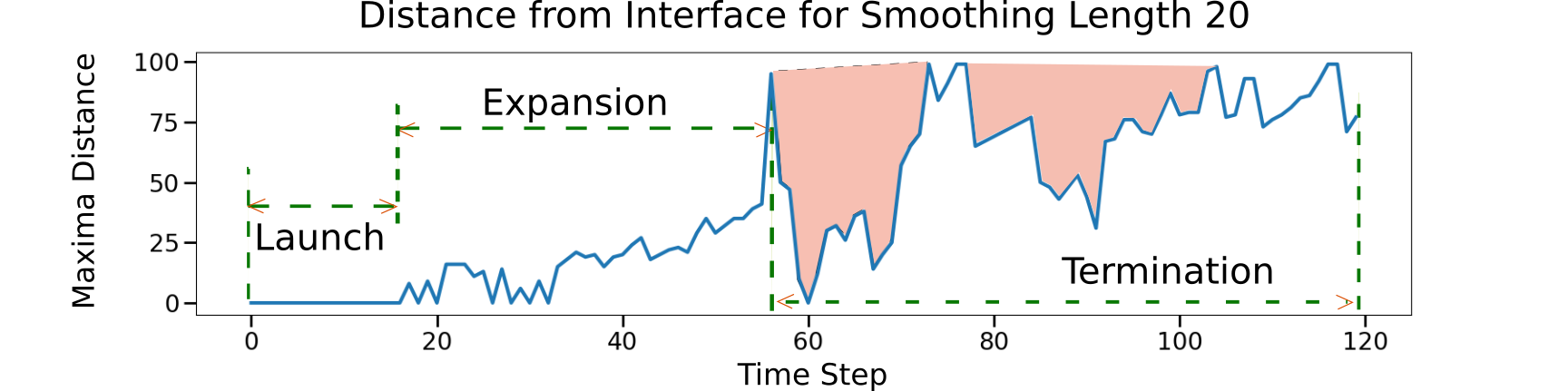}
\label{fig_vfphase_plot_sl20}}
~
\subfigure[Finger formation dynamics]{\includegraphics[width=0.95\textwidth]{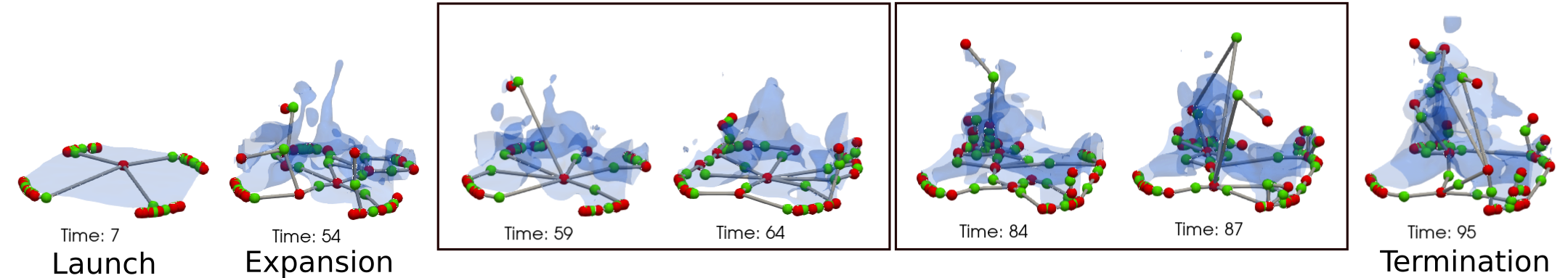}
\label{fig_vfphases}}

  \caption{Identifying viscous finger evolution phases. \subref{fig_vfphase_plot_sl20}~A plot over time of the largest distance of a maximum in the extremum graphs from the fluid mixing interface. The largest admissible value of the distance is 101, which corresponds to the boundary face opposite to the mixing interface. The plot is initially flat (\textsc{launch}), increases towards the largest admissible value (\textsc{expansion}) and then exhibits alternating steep decline and rise (\textsc{termination}). \subref{fig_vfphases}~Fingers and embedded extremum graphs at select time steps from the three phases. The box highlight shows two time steps each from the shaded regions in the \textsc{termination} phase.} 

  \label{fig_vf_regimes}
\vspace{-0.3in}
\end{figure*}

\noindent\textbf{Phases of finger evolution.}
The finger evolution can be categorized into three phases --  \textsc{launch}, \textsc{expansion}, and \textsc{termination}~\cite{favelier2016visualizing}. During the \textsc{launch} phase, the mixing interface remains almost flat and is filled with very small finger like structures. With gradual increase in the injection force, the mixing interface expands, the fingers grow larger, and the finger tip moves away from the interface during the subsequent \textsc{expansion} phase. The mixing is complete in the \textsc{termination} phase resulting in the disappearance of the fingers. Figure~\ref{fig_vfphases} shows three time steps, one from each phase, to provide an overview of finger formation. The fingers are visualized as isosurfaces at isovalue 35~\cite{lukasczyk2017,lukasczyk2019}. 

We measure the distance of all maxima from the mixing interface, defined as one of the boundary faces lying in the $z=101$ plane, and plot the largest distance over time in Figure~\ref{fig_vfphase_plot_sl20}. This plot helps identify the three phases. The initial flat region in the plot, where the distance is close to zero, is the \textsc{launch} phase. The distances increase during the subsequent \textsc{expansion} phase. The end of this phase is marked by the time step when the distance reaches the highest possible value, namely when a maximum reaches the domain boundary face opposite to the mixing interface. The following time steps belong to the \textsc{termination} phase. A finger does not evolve further once it reaches the opposite boundary face. This situation arises often towards the end of the simulation and multiple fingers terminate during this phase.

We observe two regions, shaded in Figure~\ref{fig_vfphase_plot_sl20}, on the line plot characterized by a steep decline in distances followed by gradual rise. These regions correspond to a phase consisting of dynamic changes in the fingers, where the possibility of observing topological changes due to merges and splits between components of fingers is high. Such dynamic changes can be observed both within the fingers and in the extremum graphs in time intervals $[59,64]$ and $[84,87]$, see box highlights in Figure~\ref{fig_vfphases}.

\begin{figure*}

 \centering 
 
 \subfigure[Temporal tracks for viscous fingers.]{\includegraphics[width=0.98\textwidth]{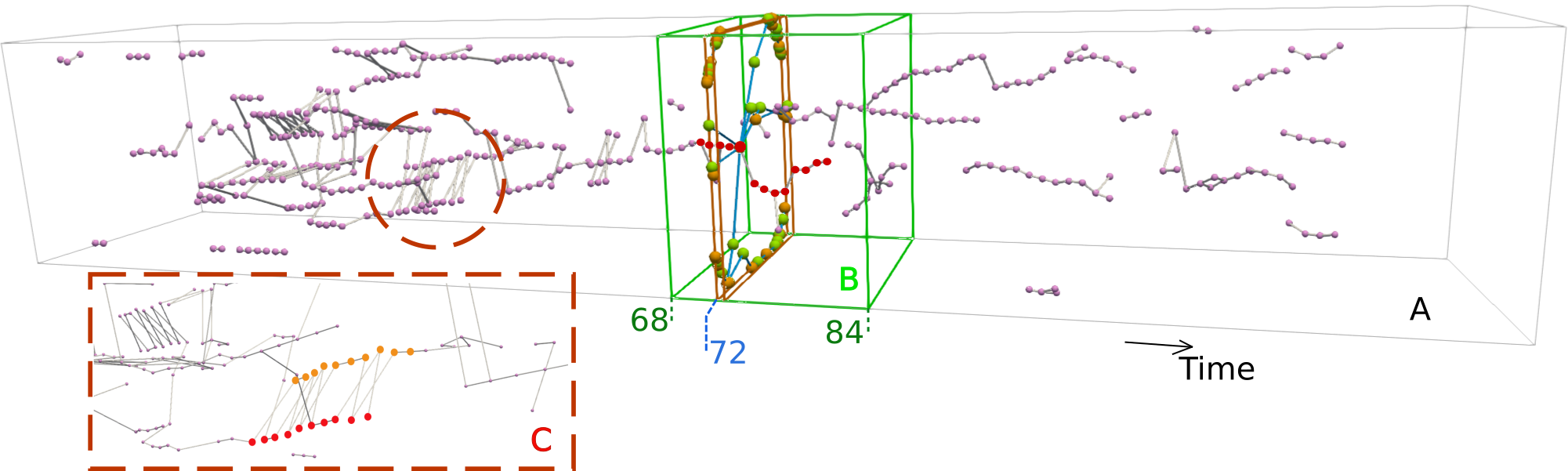}
\label{fig_vf_tracks}}
~
\subfigure[Three maxima sampled from a selected temporal track in block B.]{\includegraphics[width=0.98\textwidth]{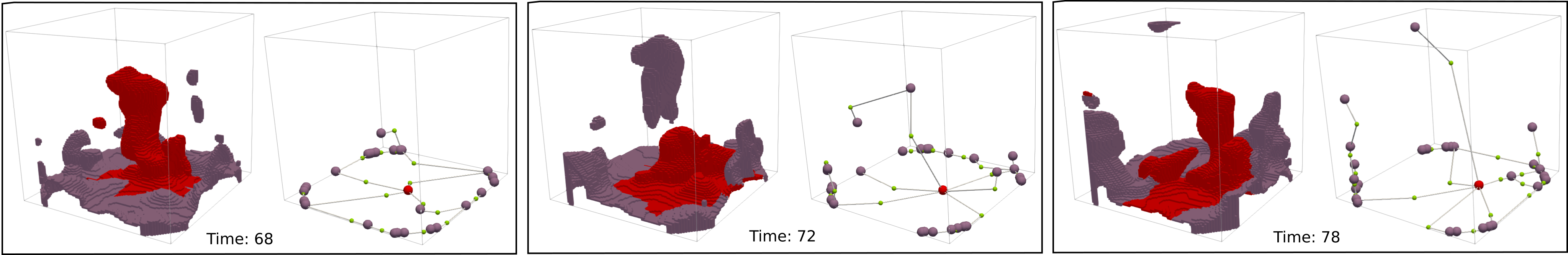}
\label{fig_vf_iso_tracks}}
~
\subfigure[Three maxima pairs sampled from two selected temporal tracks in block C.]{\includegraphics[width=0.98\textwidth]{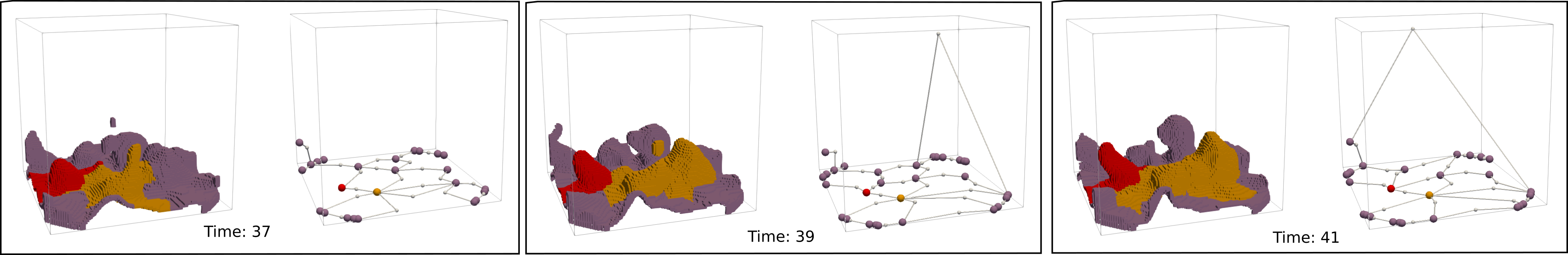}
\label{fig_vf_iso_par_tracks}}

  \caption{Temporal tracks from \TVEG for visualizing the viscous fingers dataset. \subref{fig_vf_tracks}~The 3D domain is scaled along the z-axis and stacked to visualize the tracks over time. The tracks indicate typical growing behavior of fingers along the z-axis. One track is selected (red) within a time interval [68,84] of interest (green bounding box B). A spatial region of interest with pair of highlighted tracks is shown in block C. \subref{fig_vf_iso_tracks}~Visualizing the dynamics of fingers corresponding to the selected track. The finger in red corresponds to maxima from the selected track. A split event results in a change in the shape of the red finger. \subref{fig_vf_iso_par_tracks}~Visualizing a pair of similar features along a pair of tracks highlighted in red and orange in block C. These pair of features corresponds to the three maxima pairs sampled from the red and orange sets of maxima.} 

  \label{fig_vf_tveg_tracks}
\vspace{-0.3in}
\end{figure*}

\noindent\textbf{Finger dynamics along \TVEG tracks.}
\TVEG temporal arcs can be used to support visual observation of the finger formation dynamics. We follow the method described in Section~\ref{sec_gaussians} to construct temporal arcs and visualize tracks (see Figure~\ref{fig_vf_tveg_tracks}). The domain is scaled along the z-axis. Figure~\ref{fig_vf_tracks} shows the domain and extremum graph for time step 72. The temporal tracks for all maxima lying close to the centre of the domain are displayed. Tracks of maxima near the domain boundary are not shown to reduce clutter. 

We focus on a time interval $[68,84]$, highlighted by the green bounding box, to showcase the utility of the tracks. One track (red) that appears to have a distinct behavior is selected within this time interval and studied by visualizing the associated fingers. The finger region associated with a maximum is computed as the intersection of its descending 3-manifold with the volume enclosed by the isosurface at scalar value 35. We pick three time steps (68, 72, and 78) within the time interval. Figure~\ref{fig_vf_iso_tracks} shows the extremum graph and finger regions associated with the maxima. The red finger regions are associated with maxima from the selected track. We notice a split event resulting in a change in the shape of the red region. \TVEG can thus help in visualizing finger evolution dynamics along a chosen track and help assess the contribution to finger dynamics from individual maxima in the track. 

Next, we focus on a pair of \TVEG features, highlighted in block \textbf{C} of Figure~\ref{fig_vf_tracks}. The similarity between the pair of red and orange maxima sets is characterized by a pair of \TVEG tracks that alternate between these two sets in the time interval $[34,42]$. Furthermore, the temporal dynamics of this pair of \TVEG tracks approximately mirror each other. Figure~\ref{fig_vf_iso_par_tracks} shows three maxima pairs at time instances 37, 39, and 41 and their associated finger regions. We observe that the associated finger regions form a pair of features that maintain spatial proximity and represent similar dynamic changes in finger formation along the pair of \TVEG tracks.

\noindent\majorrevision{\textbf{Relative importance of score components.}
We present a study to assess the importance of the individual components of the score that determines the correspondence, and hence the temporal arcs~(Equation~\ref{eqn:top-two-correspondences}). The components are categorized into two groups, global and local. Persistence is a global topological descriptor determined by the maxima and hence belongs to the global group. The local group consists of three score components that represent local features: function value, distance between maxima, and the neighborhood similarity between maxima. In this study, we vary the relative importance of the score components by tuning their corresponding weights. We present a qualitative analysis based on comparison of the \TVEG tracks computed for each assignment of  weights. To explain the weight settings, let us adopt the notation $(G,L)$, $G+L=1$, to indicate a weight $G$ assigned to the global component and $L$ assigned to the local components. The weight $L$ is uniformly divided amongst all three components of the local group. The \TVEG tracks are computed for five such weight settings: $(0.25,0.75), (0.5,0.5), (0.75,0.25), (1,0),$ and $(0,1)$. Figure~\ref{fig_scw_comparison} shows the computed tracks for a qualitative analysis. 
\begin{figure*}

\centering
   
\subfigure[$(0.25,0.75)$]{\includegraphics[width=0.3\textwidth]{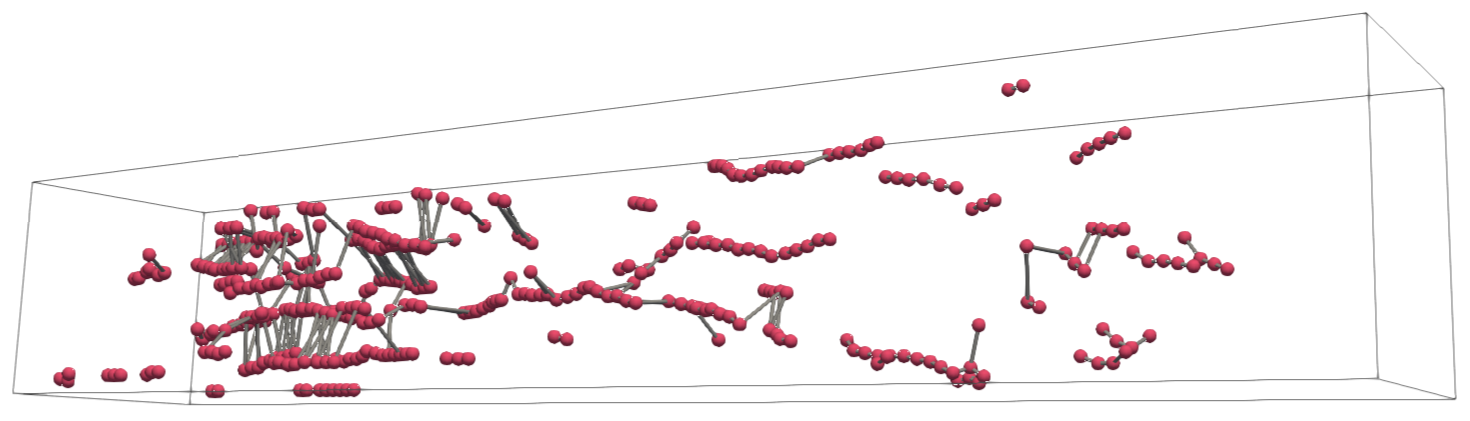}
\label{fig_scw_1}}
~
\subfigure[$(0.5,0.5)$]{\includegraphics[width=0.3\textwidth]{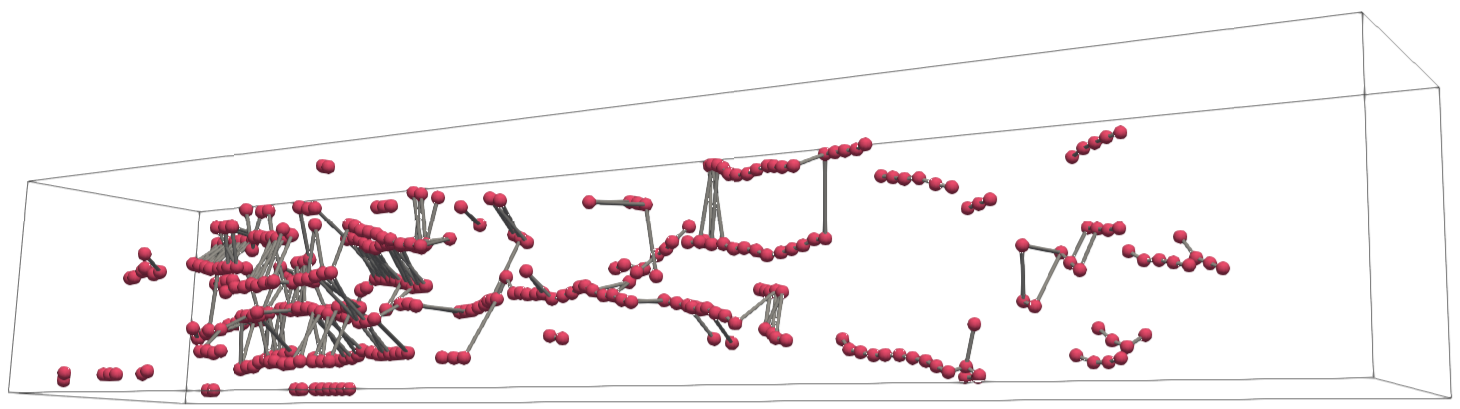}
\label{fig_scw_4}}
~
\subfigure[$(0.75,0.25)$]{\includegraphics[width=0.3\textwidth]{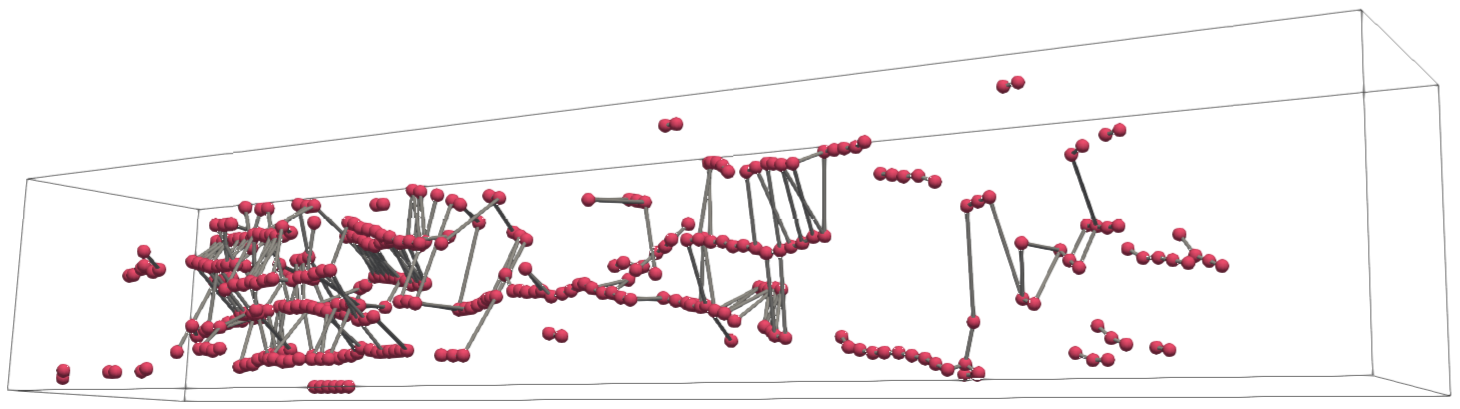}
\label{fig_scw_3}}
~
\subfigure[$(1,0)$]{\includegraphics[width=0.3\textwidth]{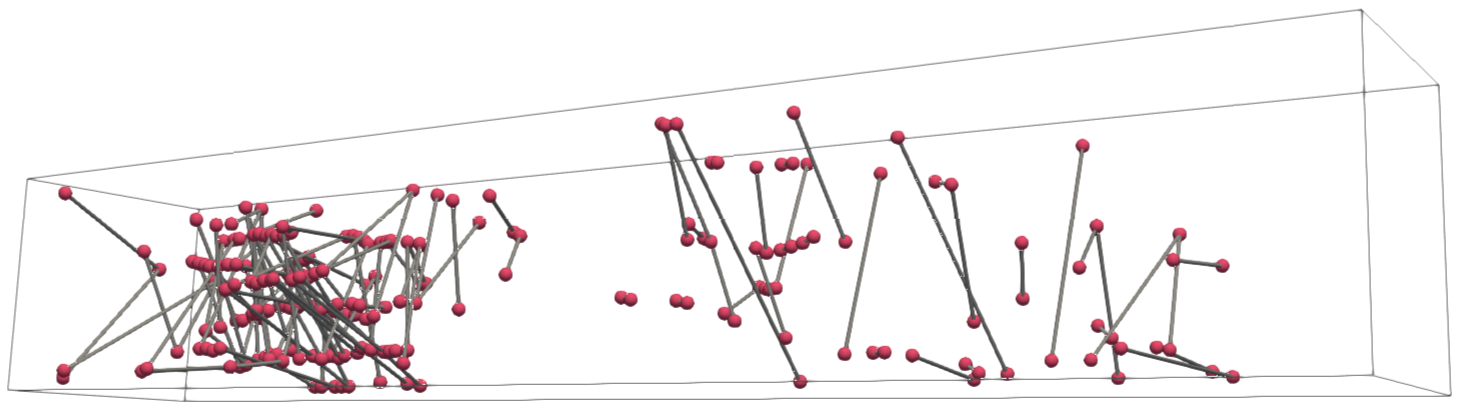}
\label{fig_scw_2}}
~
\subfigure[$(0,1)$]{\includegraphics[width=0.3\textwidth]{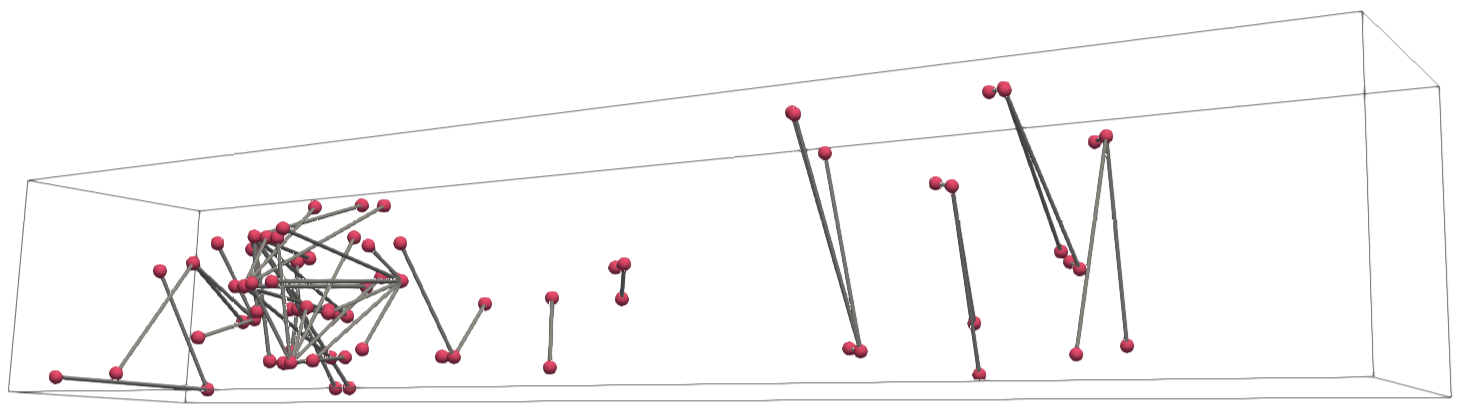}
\label{fig_scw_5}}

\caption{\majorrevision{Studying the effect of score component weights on viscous fingers tracks. Weights of global component persistence and combined weights of all local components are shown separately.  \TVEG tracks fail to explain finger formation dynamics when either local~(d) or global component~(e) is ignored completely, as suggested by large gaps within the computed tracks. On the other hand, assigning non-zero weights to both local and global components~(a, b, c) helps capture the tracks that are important to explain finger formations.}}

\label{fig_scw_comparison}

\end{figure*}

Figures~\ref{fig_scw_1},~\ref{fig_scw_4}, and~\ref{fig_scw_3} represent the \TVEG tracks computed under weight assignment $(0.25,0.75), (0.5,0.5)$ and $(0.75,0.25)$, respectively. Here, both global and local components were incorporated by assigning them a non-zero weight. Figure~\ref{fig_scw_2} and~\ref{fig_scw_5} represents settings $(1,0)$ and $(0,1)$, respectively, which exclusively considered either the global or local components. Both Figure~\ref{fig_scw_2} and~\ref{fig_scw_5} can be characterized as consisting of a significant number of small \TVEG tracks with large gaps between the tracks. On the other hand, \TVEG tracks in Figures~\ref{fig_scw_1},~\ref{fig_scw_4}, and~\ref{fig_scw_3} are spatially consistent and relatively longer, which corresponds to the gradual formation of fingers. Furthermore, Figures~\ref{fig_scw_4} and~\ref{fig_scw_3}, where the global persistence component is assigned a relatively higher weight, shows more number of abrupt jumps in the \TVEG tracks in comparison to tracks obtained by assigning equal weights \minorrevision{to all four components \ie $(G,L)=(0.25,0.75)$ (Figure~\ref{fig_scw_1}). Results from an additional experimental study of the effect of weight assignments on the computed arcs is available in the supplementary material. We also performed additional tests at finer resolution, where the $(G,L)$ parameters were varied with smaller step sizes, specifically vary $G$ by $\epsilon = \pm 0.01$. However, no significant improvement was observed as compared to a weight assignment of $(G,L)=(0.25,0.75)$ in all datasets. All experiments relevant to the case studies in Section~\ref{sec_case_studies} are performed with a weight assignment $(G,L)=(0.25,0.75)$.}

These experiments show that discarding the local score components results in irregular short \TVEG tracks. Relying only upon local components also results in tracks that lack meaningful consistency over time. A combination of local and global components, preferably weighted uniformly, can lead to more meaningful \TVEG tracks capable of assisting effective data dynamics visualization.}

\subsection{Vortex street}\label{sec_cs_vortex}
\begin{figure*}

\centering
 
\subfigure{\includegraphics[width=0.18\textwidth]{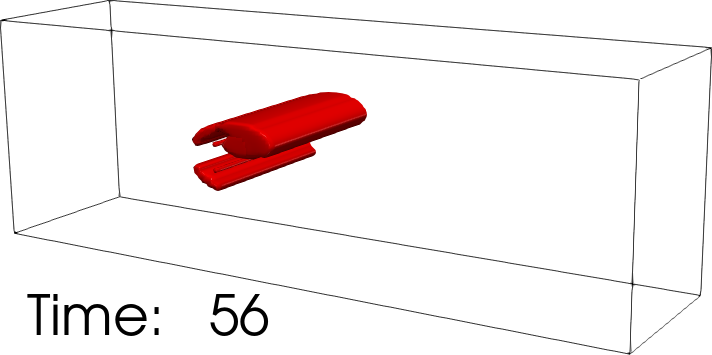}
\label{fig:t56}}
~
\subfigure{\includegraphics[width=0.18\textwidth]{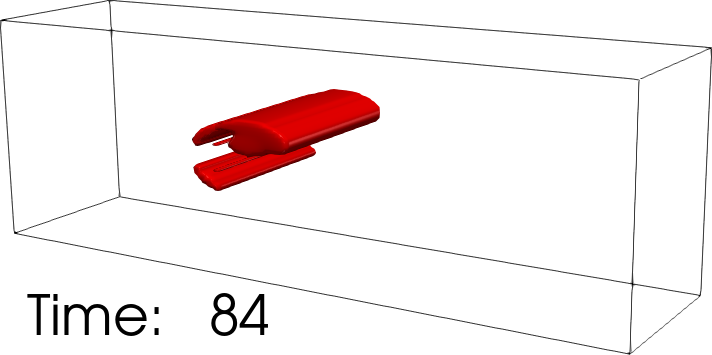}
\label{fig:t84}}
~
\subfigure{\includegraphics[width=0.18\textwidth]{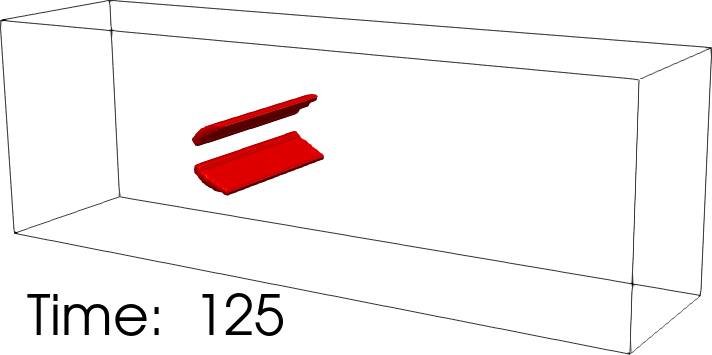}
\label{fig:t125}}
~
\subfigure{\includegraphics[width=0.18\textwidth]{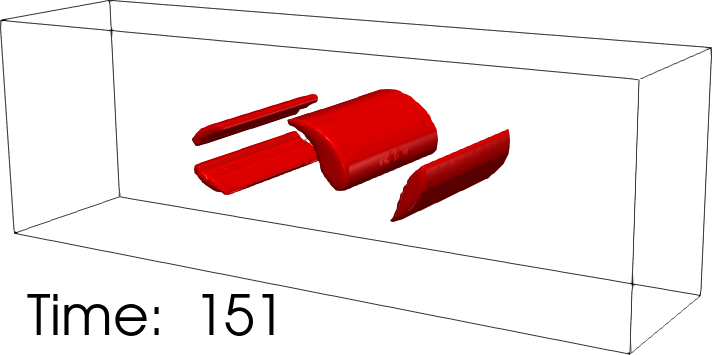}
\label{fig:t151}}
~
\subfigure{\includegraphics[width=0.18\textwidth]{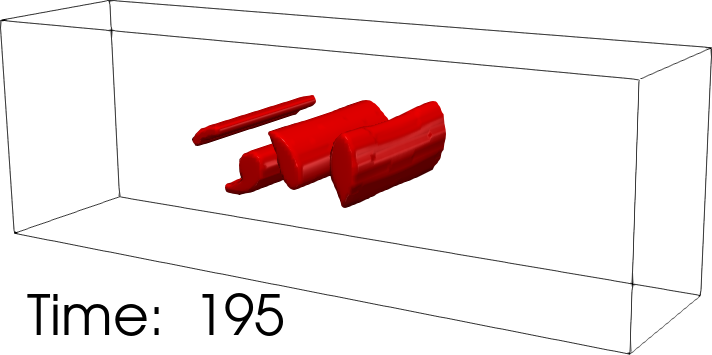}
\label{fig:t195}}
~
\subfigure{\includegraphics[width=0.18\textwidth]{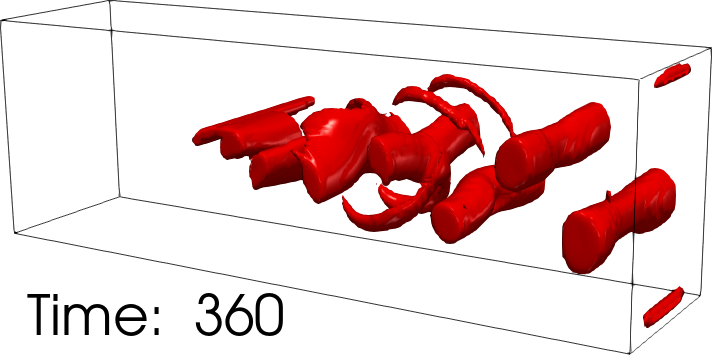}
\label{fig:t360}}
~
\subfigure{\includegraphics[width=0.18\textwidth]{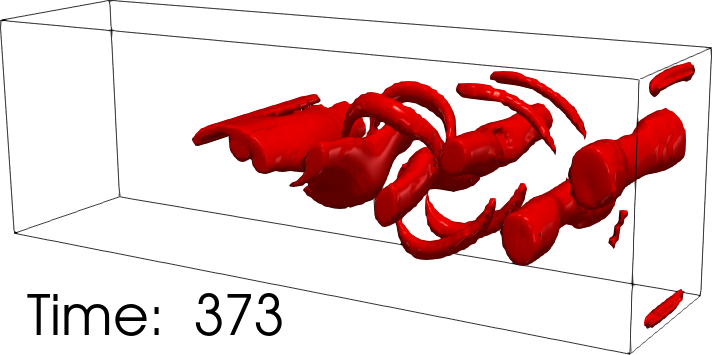}
\label{fig:t373}}
~
\subfigure{\includegraphics[width=0.18\textwidth]{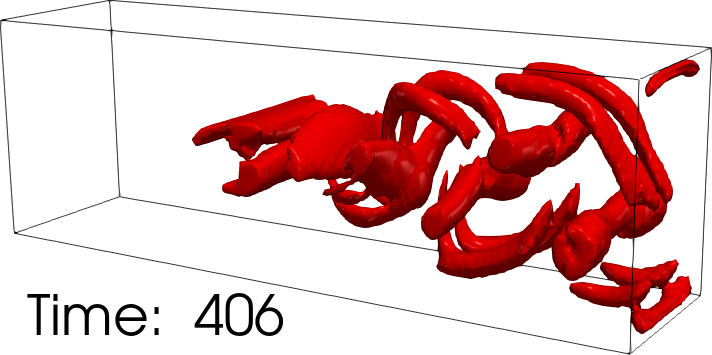}
\label{fig:t406}}
~
\subfigure{\includegraphics[width=0.18\textwidth]{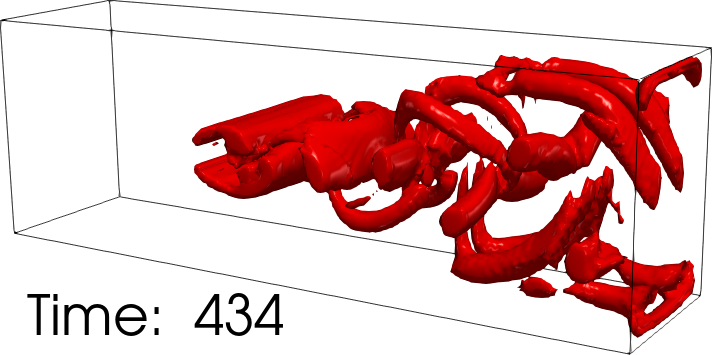}
\label{fig:t434}}
~
\subfigure{\includegraphics[width=0.18\textwidth]{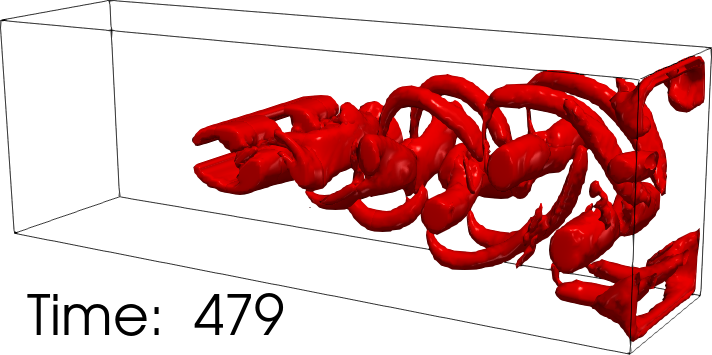}
\label{fig:t479}}

\caption{Visualizing the top $20\%$ tracks in the 3D von K\'arm\'an vortex street data in terms of track length. The tracks are generated from the \TVEG based on spatial overlaps. Regions associated with maxima in the tracks are displayed. The top tracks include the temporal evolution of primary and secondary vortices.}

\label{fig:total}

\end{figure*}

\begin{figure*}

\centering
 
\subfigure{\includegraphics[width=0.18\textwidth]{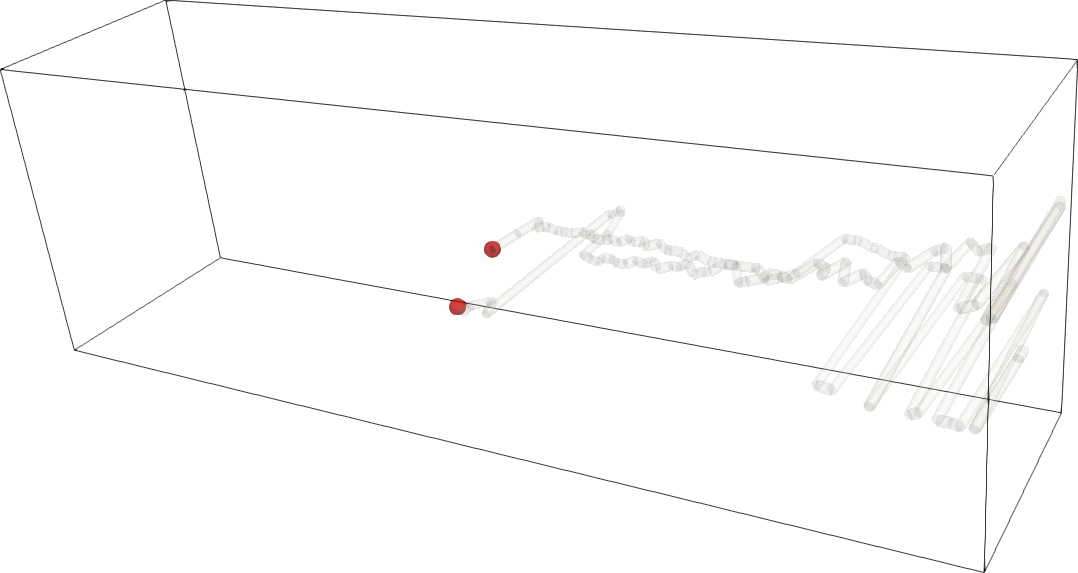}
\label{fig:q1}}
~
\subfigure{\includegraphics[width=0.18\textwidth]{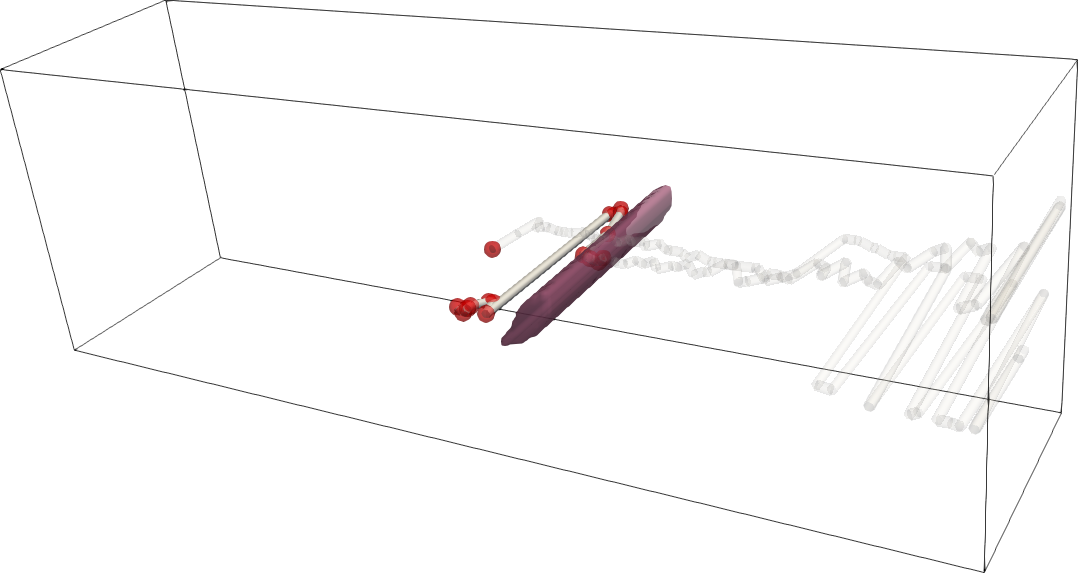}
\label{fig:q2}}
~
\subfigure{\includegraphics[width=0.18\textwidth]{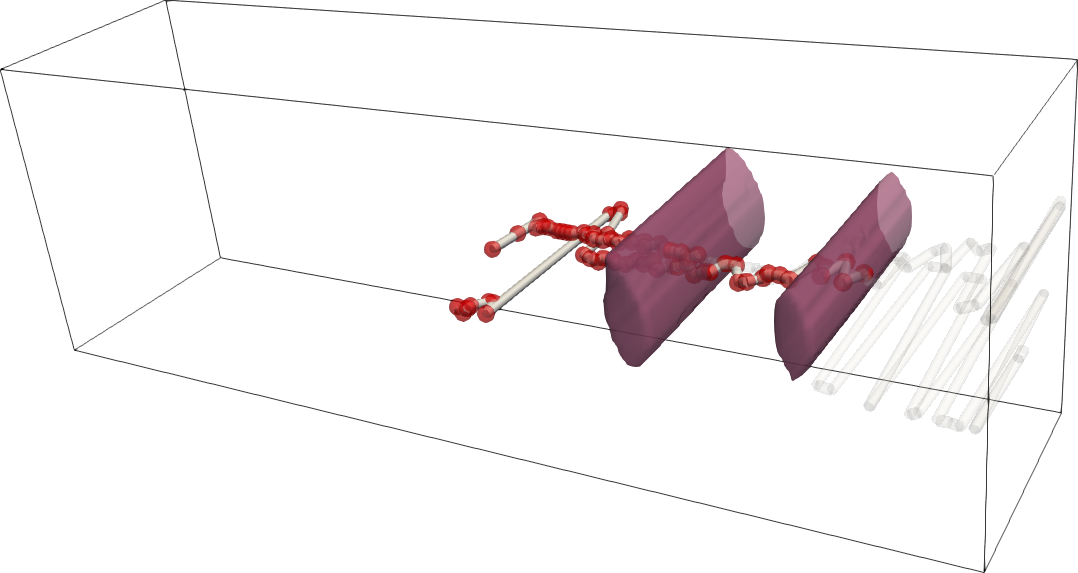}
\label{fig:q3}}
~
\subfigure{\includegraphics[width=0.18\textwidth]{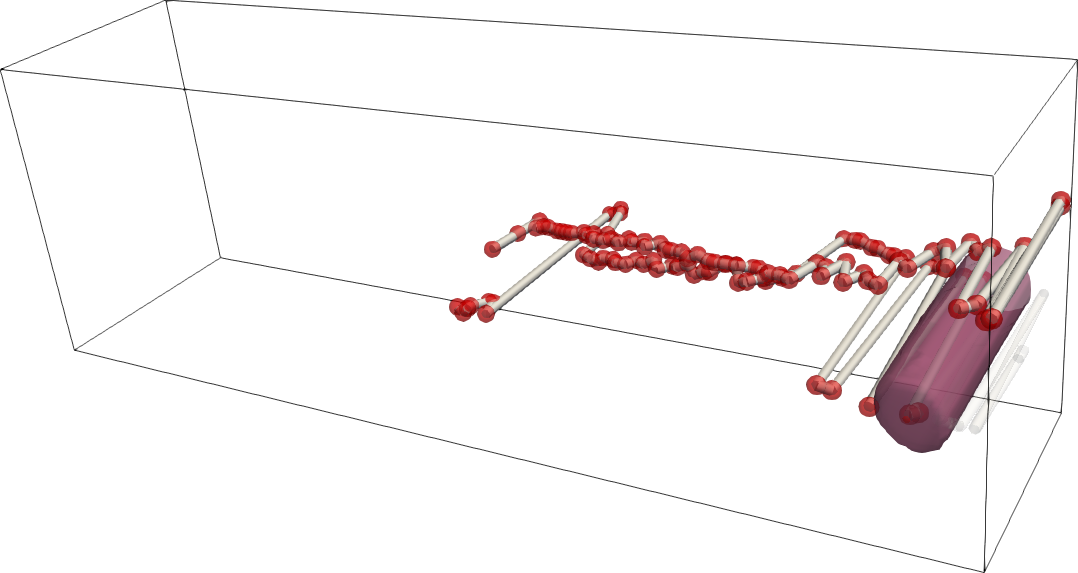}
\label{fig:q4}}
~
\subfigure{\includegraphics[width=0.18\textwidth]{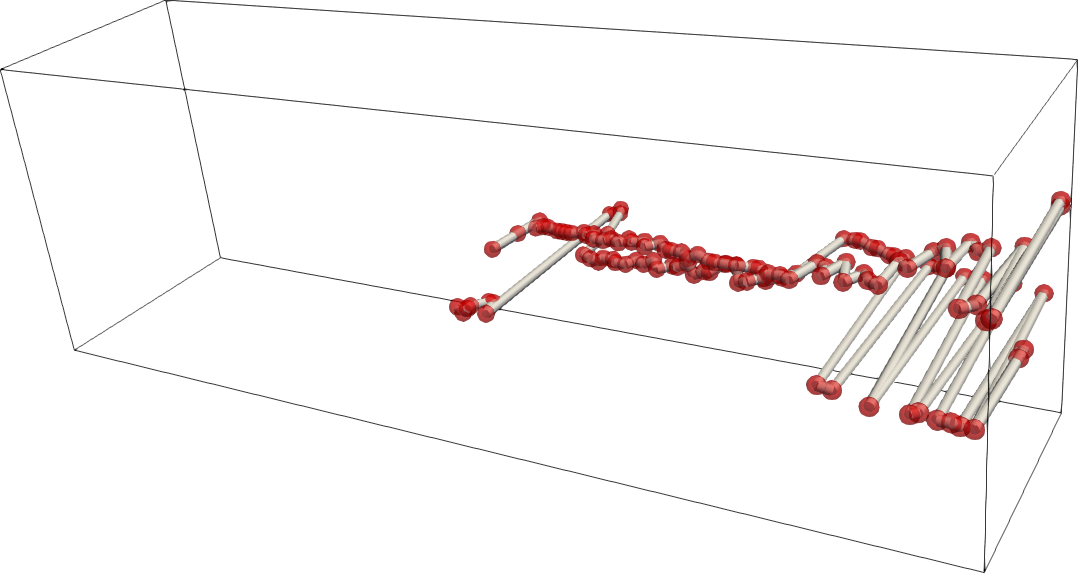}
\label{fig:q5}}
~
\caption{\majorrevision{Visualizing tracks computed in response to a spatiotemporal query regarding two maxima}~(left). The spatial location of the maxima varies substantially even though they represent the same feature over time. (left to right)~As the maxima and corresponding features evolve over time, the opacity of the track is increased and the maxima are also displayed.}

\label{fig:query}
\vspace{-0.3in}
\end{figure*}

\begin{figure}[ht]
\centering
\subfigure[Time step $223$]{\includegraphics[width=0.14\textwidth]{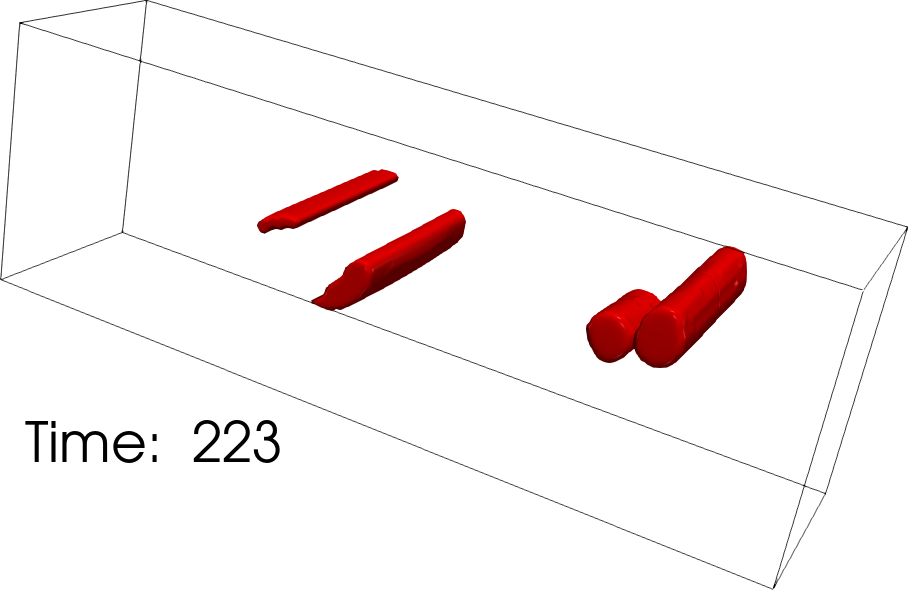}
\label{fig:t223}}
~
\subfigure[Time step $224$]{\includegraphics[width=0.14\textwidth]{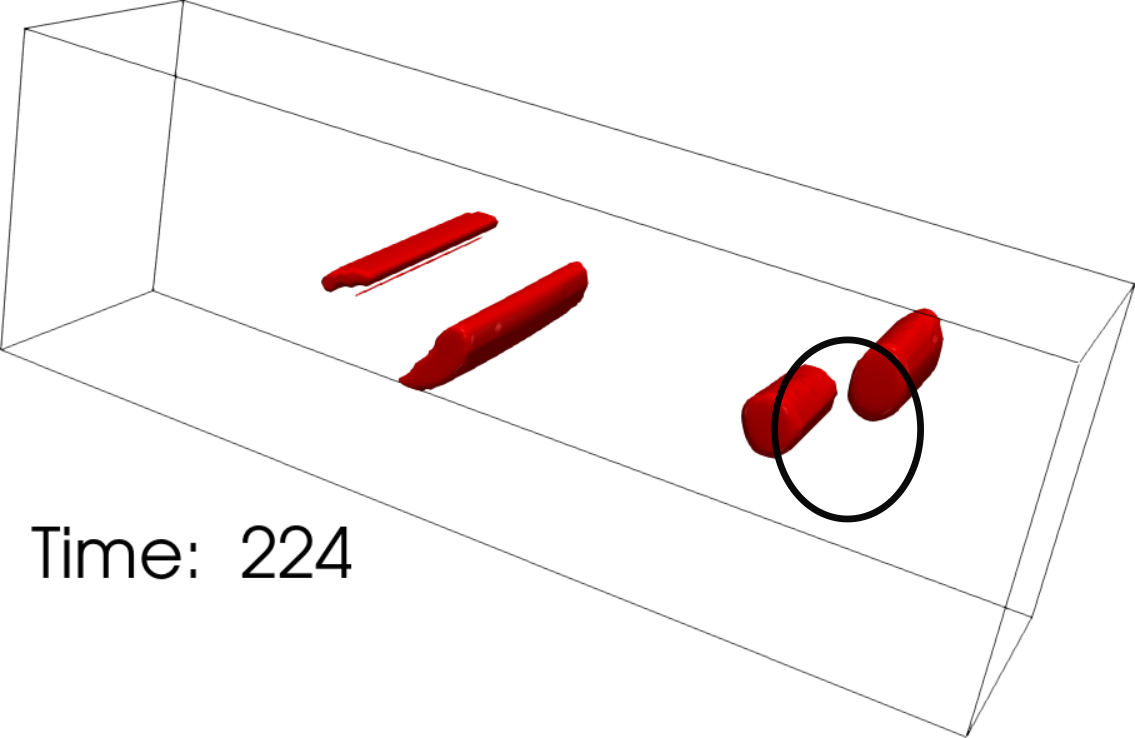}
\label{fig:t224}}
~
\subfigure[Time step $225$]{\includegraphics[width=0.14\textwidth]{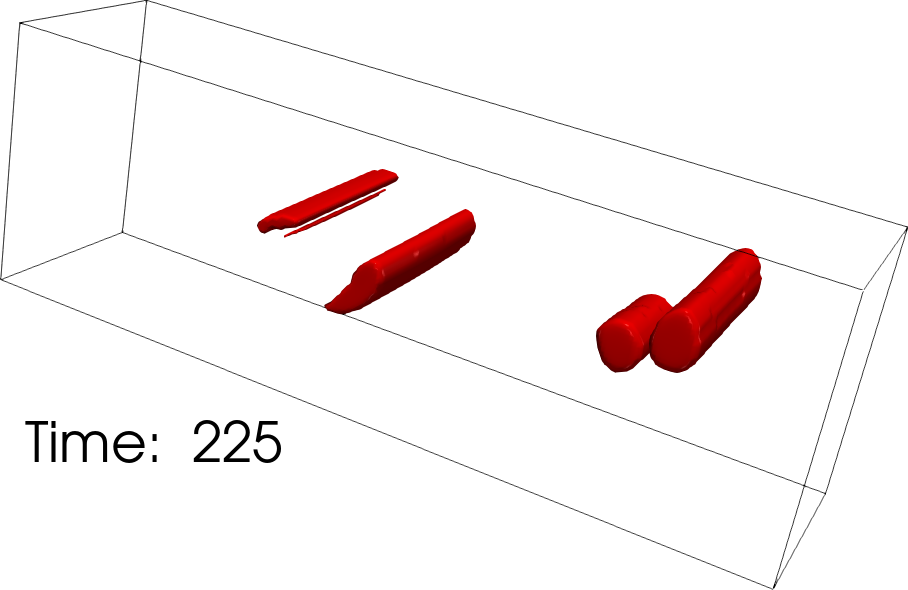}
\label{fig:t225}}

\subfigure[Temporal arcs between $224$ (left) and $225$ (right)] {\includegraphics[width=0.30\textwidth]{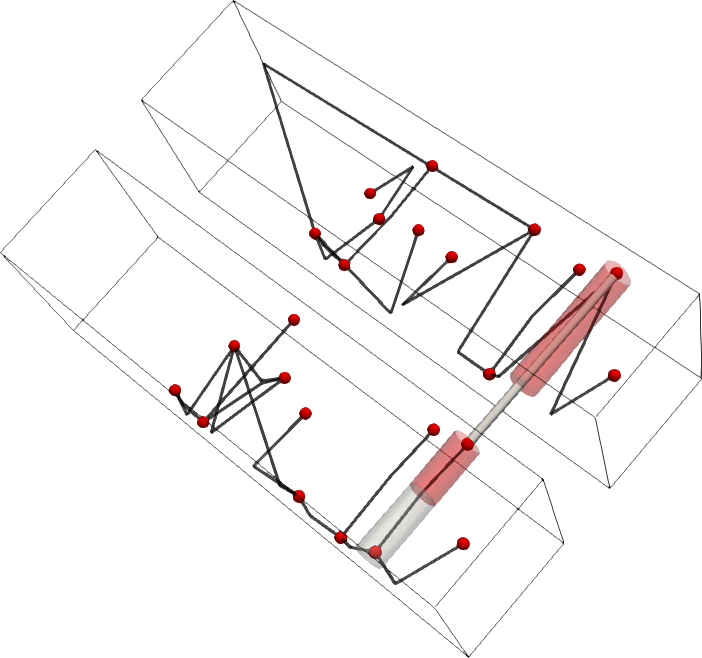}
\label{fig:t224_225}}

\caption[Illustrating challenges in tracking vortices in the 3D vortex street]{Illustrating challenges in tracking vortices in the 3D von K\'arm\'an vortex street data. \subref{fig:t223}-\subref{fig:t225}~Vortices in time steps 223-225. \subref{fig:t224_225}~Temporal arc (gray, bold) between time steps 224-225 for a particular feature and the extremum graph arcs (black). The white feature in time 224 should have ideally been mapped to the red feature in time 225. But, no temporal arc is incident on it. The red features in 224 and 225 are connected by a temporal arc.}

\label{fig:except}
\vspace{-0.3in}
\end{figure}

We demonstrate the utility of \TVEG in feature tracking by presenting two use cases -- computing a summary view by tracking features that are automatically computed, and tracking features  that are specified as a \majorrevision{spatiotemporal user query}. We demonstrate both use cases on a 3D B\'enard-von K\'arm\'an vortex street dataset. The Okubo-Weiss criterion is a scalar field that indicates regions of high vorticity~\cite{saikia2017}. It is sampled on a $192 \times 64 \times 48$ regular grid over $508$ time steps. Previous results~\cite{saikia2017,Sridh2021} identify two classes of vortices, primary and secondary, and track them across time. 

\noindent \textbf{Overview of features.} 
In order to identify and track all topological features that can be captured using an extremum graph, we first compute the \TVEG and consider all temporal arcs. Each maximum has an associated region, which is spanned by its descending manifold clipped by the isosurface at scalar value 0.1. The threshold value of 0.1 represents high vorticity regions.

We calculate spatial overlaps between the regions associated with a maximum in time step $t$ and its two correspondences in time step $t+1$, and pick the temporal arc with the larger overlap. Arcs where the spatial overlap is small and short tracks (whose length is less than 10 time steps) are removed. We observe that each secondary vortex is represented by a single maximum, but a primary vortex may comprise of regions associated with a collection of maxima. In the latter case, the multiple maxima spanning a primary vortex are connected in the extremum graph via saddles. So, we collate such tracks in a post-processing step. We observe that the top $\approx 20\%$ tracks sorted in decreasing order of track length consists of many of the primary and secondary vortices, see Figure~\ref{fig:total}. The tracks are visualized by displaying the regions associated with maxima at different time steps. We observe that none of the temporal arcs connect a primary vortex with a secondary. So, the \TVEG provides a good summary of the two types of vortices.

\noindent \textbf{Query driven feature tracking.} 
User queries consisting of a set of maxima from primary or secondary vortices or both may be used to visualize specific tracks. Figure~\ref{fig:query} shows one such query. Again, we observe that the temporal arcs either connect two primary vortices or two secondary vortices. One challenge that is typically encountered while analyzing this vortex street data is that, even though the spatial movement of vortices is along a smooth curve, the maxima that represents the vortices follow a circuitous path. The temporal arcs in the \TVEG play a crucial role in finding the correspondences between such maxima. It provided a foundation upon which the additional spatial overlap criterion could be applied. The videos in the supplementary material accompanying the paper demonstrate both use cases, namely visualization of top tracks and tracks specified by user queries.

\noindent \textbf{Discussion.} 
Overall, we find that the \TVEG serves as a good underlying representation that supports the development of feature tracking methods. The four components of the objective function are sufficient to identify meaningful correspondences in most cases, but there are exceptions. 

The simplification threshold which is uniformly applied across all time steps affects the extremum graph and therefore the segmentation. While a maximum may continue to represent the same vortex across different time steps, its spatial location within the vortex may vary substantially as seen in Figure~\ref{fig:query}. These two factors may lead to instances where the top two temporal arcs from a maximum in time $t$ to maxima in time $t+1$ may not represent the desired feature correspondences. Further, filtering the correspondences in a refinement step may result in zero temporal arcs incident on a maximum and hence discontinuation of tracks.%(\ie tracks with zero length). 

We show one such case occurring in time steps 223-225 in Figure~\ref{fig:except}. A feature represented by a maximum in 223 splits, resulting in a primary vortex represented by two maxima. One of those maxima, which represents a partial vortex in time 224, contains temporal arcs to features in time 225, but there is no arc for one feature (see Figure~\ref{fig:t224_225}). As a consequence, the track containing the white feature in time 224 terminates in time 225, thereby causing the anomaly highlighted in Figure~\ref{fig:t224}. 
Such early terminations lead to smaller length tracks in the 3D vortex street, specifically for primary vortices. This results in a smaller number of primary vortices among the top tracks as shown in the supplemental video.

The neighborhood component $\mathcal{N}$ of the score helps balance the effect of the variation in the location of maxima. A time step-specific simplification threshold may have resulted in a single maximum representing the primary vortex, alleviating this anomaly. 

%A temporal arc may identify correspondence between two features in such a way that spatially they might be oriented in a direction different from the direction of the feature movement. In the vortex street this might lead to a temporal arc between one vortex and another vortex which follows the first one. Such scenarios are data specific, and require application knowledge for appropriate handling. 
%\vijay{This is a hypothetical situation with a vague description. Do we want to include it?}
%\raghavendra{I remember in the earlier version without the neighborhood information, this was occurring, so I retained the text, we can exclude it.}

%% file: section5.tex
\section{Conclusions}\label{Conclusions}
We introduced \TVEG, a time-varying extremum graph to facilitate analysis and visualization of time-varying scalar fields. To the best of our knowledge, \TVEG is the first time-varying topological structure based on extremum graphs. The structure is easy to compute and interpret. We demonstrate its utility in feature tracking tasks and for analysis of synthetic and simulated data. The criteria for computing temporal correspondences may be extended to incorporate other attributes of critical points and extremum graphs, \majorrevision{and to support  multi-way split and merge events. The representation capabilities and queries supported by \TVEG help in the study of local and global topological events. Extending these capabilities will facilitate the visual exploration and analysis of complex datasets.}

%% file: tveg_app_cgf.tex
% for anonymous conference submission please enter your SUBMISSION ID
% instead of the author's name (and leave the affiliation blank) !!
% for final version: please provide your *own* ORCID in the brackets following \orcid; see https://orcid.org/ for more details.
%\author[D. Fellner \& S. Behnke]
%{\parbox{\textwidth}{\centering D.\,W. Fellner\thanks{Chairman Eurographics Publications Board}$^{1,2}$\orcid{0000-0001-7756-0901}
%        and S. Behnke$^{2}$\orcid{0000-0001-5923-423X} 
%%        S. Spencer$^2$\thanks{Chairman Siggraph Publications Board}
%        }
%        \\
%% For Computer Graphics Forum: Please use the abbreviation of your first name.
%{\parbox{\textwidth}{\centering $^1$TU Darmstadt \& Fraunhofer IGD, Germany\\
%         $^2$Graz University of Technology, Institute of Computer Graphics and Knowledge Visualization, Austria
%%        $^2$ Another Department to illustrate the use in papers from authors
%%             with different affiliations
%       }
%}
%}
% ------------------------------------------------------------------------

% if the Editors-in-Chief have given you the data, you may uncomment
% the following five lines and insert it here
%
% \volume{36}   % the volume in which the issue will be published;
% \issue{1}     % the issue number of the publication
% \pStartPage{1}      % set starting page

%-------------------------------------------------------------------------
%\begin{document}
\onecolumn
% uncomment for using teaser
% \teaser{
%  \includegraphics[width=\linewidth]{eg_new}
%  \centering
%   \caption{New EG Logo}
% \label{fig:teaser}
%}

%\maketitle
%-------------------------------------------------------------------------
\newpage 

\section*{Supplementary Material for ``Time-varying Extremum Graphs''}

\section*{Abstract}
\highlightadd{This document presents additional material supporting the paper ``Time-varying Extremum Graphs''. It provides pseudo-code for the algorithm and subroutines used for computing the \TVEG, and an explanation of the method used to reduce visual clutter in the \TVEG tracks computed for the viscous finger data. Next, it presents an additional comparison of \TVEG tracks computed using varying weights assigned to the correspondence score components. It also presents detailed runtimes for \TVEG computation. \minorrevision{Finally, the major aspects of the visual analysis pipeline used in the case studies are explained.}}

%-------------------------------------------------------------------------
%\input{appendix_cgf.tex}
%\twocolumn[\begin{@twocolumnfalse}
\setcounter{section}{0}
\section{Algorithms and psuedocode}\label{sec_algorithms}
We present the pseudocode for all the algorithms related to the computation of \TVEG. \textsc{TemporalArcs} (Algorithm~\ref{algo_tem_arcs}) computes all the temporal arcs. \textsc{UpdateMergeSplitEdges} (Algorithm~\ref{algo_updateMergeSplitEdges}) updates the set of merge and split event sets after removing the highest score edge that participates in both a merge and split event. \textsc{ComputeScores} (Algorithm~\ref{algo_computeScores}) computes and returns the two best correspondences for each maximum in a time step and the associated scores. \textsc{FilterScores} (Algorithm~\ref{algo_filterscores}) refines these correspondences based on a threshold $\tau$. \textsc{DetectMerge},  \textsc{DetectSplit},  \textsc{DetectDel}, \textsc{DetectGen} (Algorithms~\ref{algo_detectmerge},\ref{algo_detectsplit},\ref{algo_detectdel},\ref{algo_detectgen}) detect topological events merge, split, deletion, and creation respectively. \textsc{ExGraph3D} (Algorithm~\ref{algo_static_eg_app}) computes the extremum graph of the time-varying scalar field at a given time step and simplifies the graph using a threshold \minoredits{$\theta$}. For the case studies that involve 3D data, we use \textsc{ExGraph3D} to compute extremum graphs, which in turn calls \textsc{MS3D}~\cite{shivshankar2012ms3dparallel,delgado2014skeletonization,bhatia2018topoms} to compute the MS complex. The extremum graph could also be directly computed from the scalar field~\cite{correa2011topological}. Table~\ref{table_cp_fields_app} contains the list of attributes of a critical point in an extremum graph.
%\end{@twocolumnfalse}
%]
%
%

\section{\highlightadd{Selection of \TVEG tracks via cropping}}\label{sec_cropping}

\begin{figure*}[!hb]
\centering
\subfigure[All \TVEG tracks]{\includegraphics[width=0.48\textwidth]{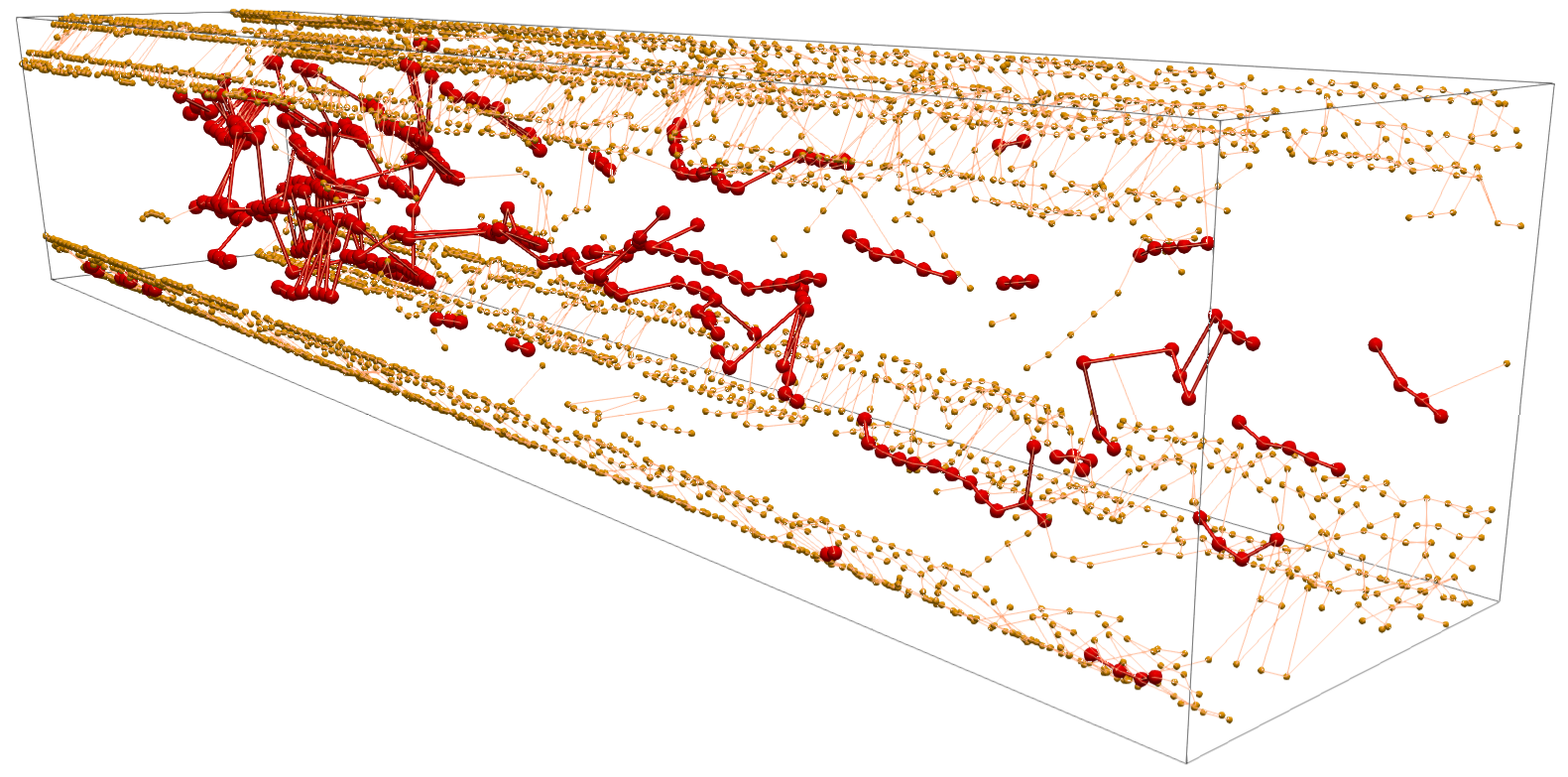}
\label{fig_app_vis_nocrop}}
~
\subfigure[Selected tracks after cropping]{\includegraphics[width=0.48\textwidth]{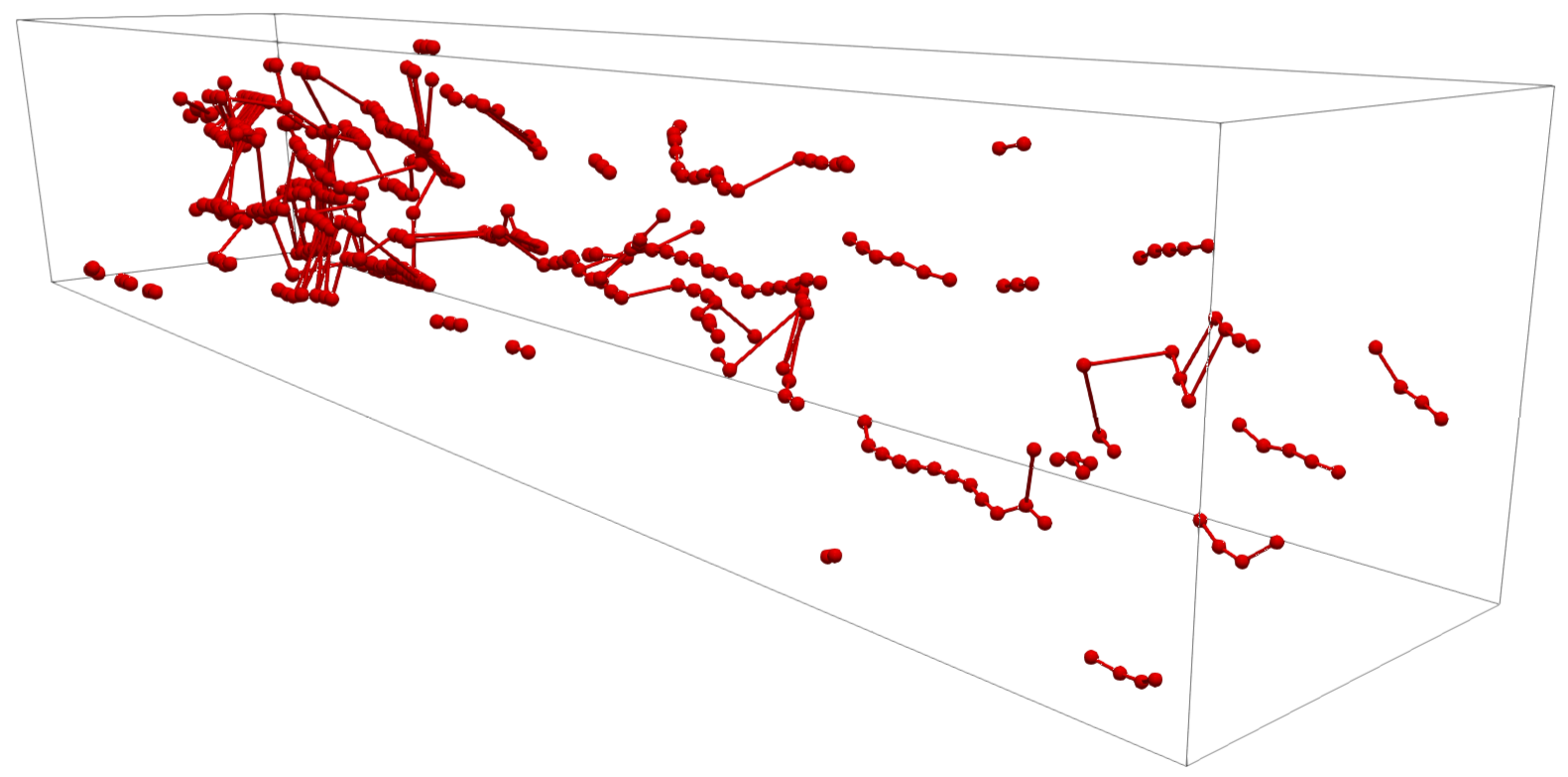}
\label{fig_app_vis_crop}}
\caption{\highlightadd{(a)~Computation of all \TVEG tracks from the viscous finger dataset . Orange tracks are located near the domain boundary. (b)~Cropping selects the red tracks, which represent the viscous finger formation.}}
\label{fig_app_crop_nocrop}
\end{figure*}

\highlightadd{\TVEG tracks for the viscous fingers dataset includes a clusters of tracks near the domain boundary. These tracks are shorter in length and clutter the visualization as shown in Figure~\ref{fig_app_vis_nocrop}. We observed that the major fingers are formed along the central part of the domain and not near the boundary. The \TVEG tracks consisting of temporal arcs that lie near the central part of the domain, shown in red, may be highlighted by removing the arcs near the boundary. Such a collection of tracks is shown in Figure~\ref{fig_app_vis_crop}. In the paper, we employ cropping to present the \TVEG tracks that play an important role in explaining the data dynamics. Cropping essentially discards temporal arcs whose endpoint maxima lie within a certain distance threshold from the domain boundary.}

\section{\highlightadd{Qualitative analysis of score components}}\label{sec_score_component}

\begin{figure*}
\centering
% \subfigure[]{\includegraphics[width=0.48\textwidth]{fig/scw_comp/appendix/common_f.png}
% \label{fig_app_scw_common}}
% ~
\subfigure[]{\includegraphics[width=0.52\textwidth]{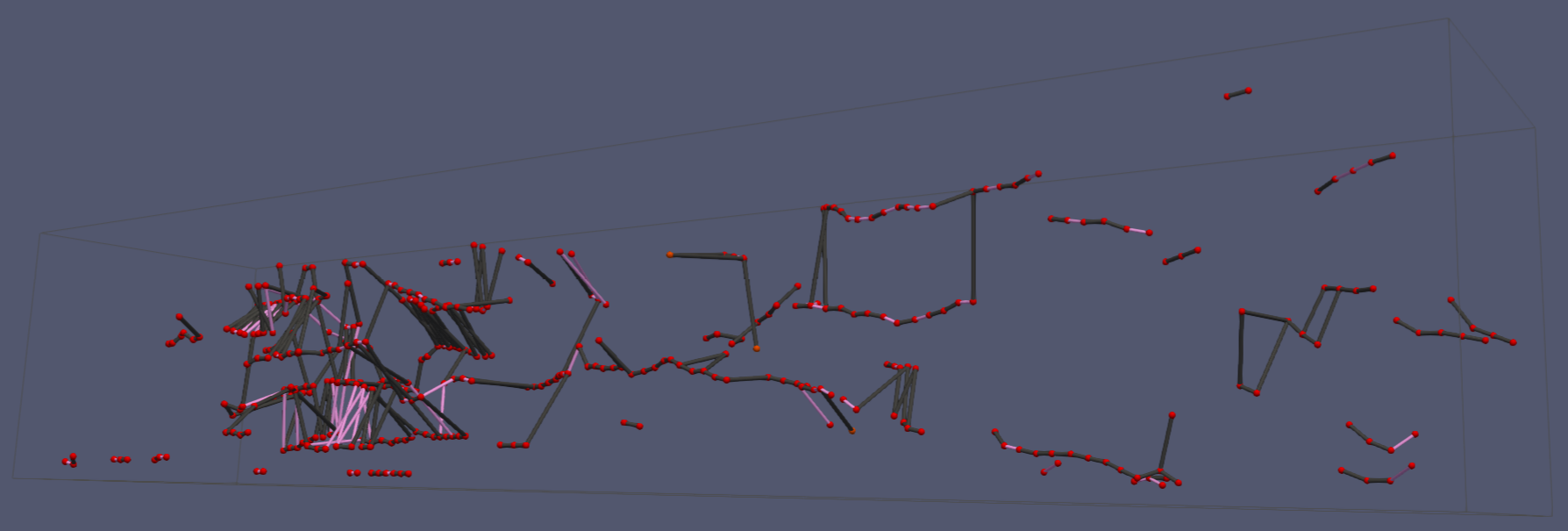}
\label{fig_app_scw_1}}
~
\subfigure[]{\includegraphics[width=0.52\textwidth]{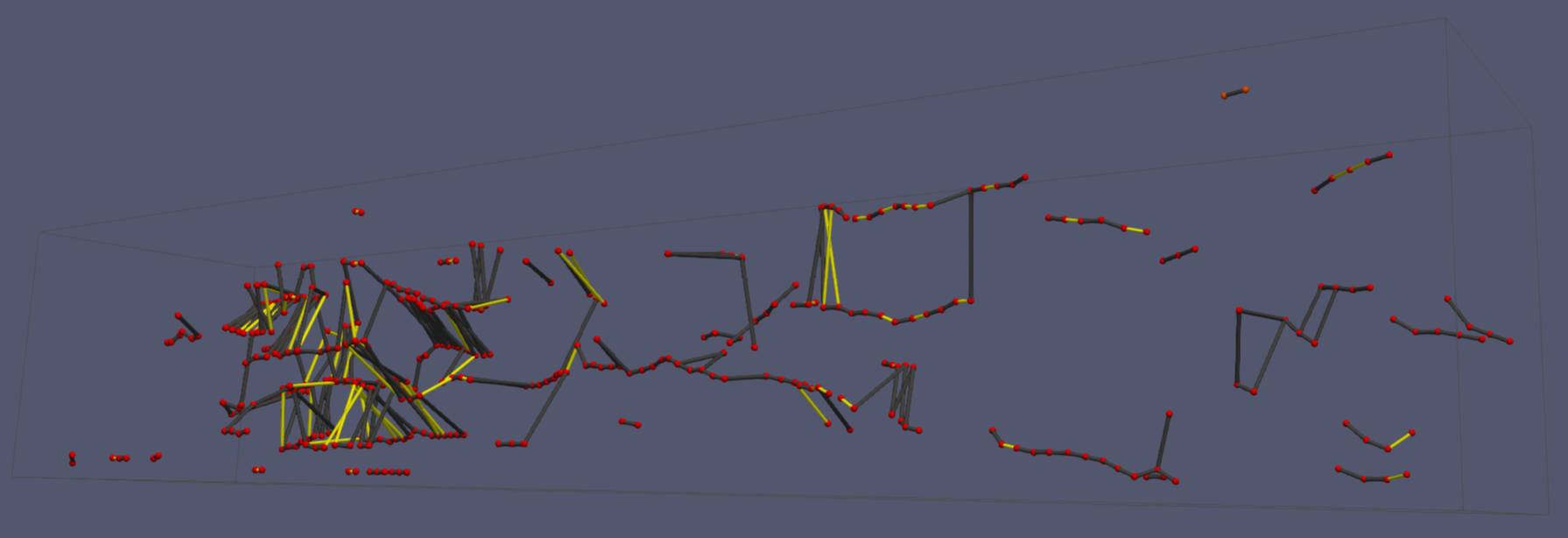}
\label{fig_app_scw_4}}
~
\subfigure[]{\includegraphics[width=0.52\textwidth]{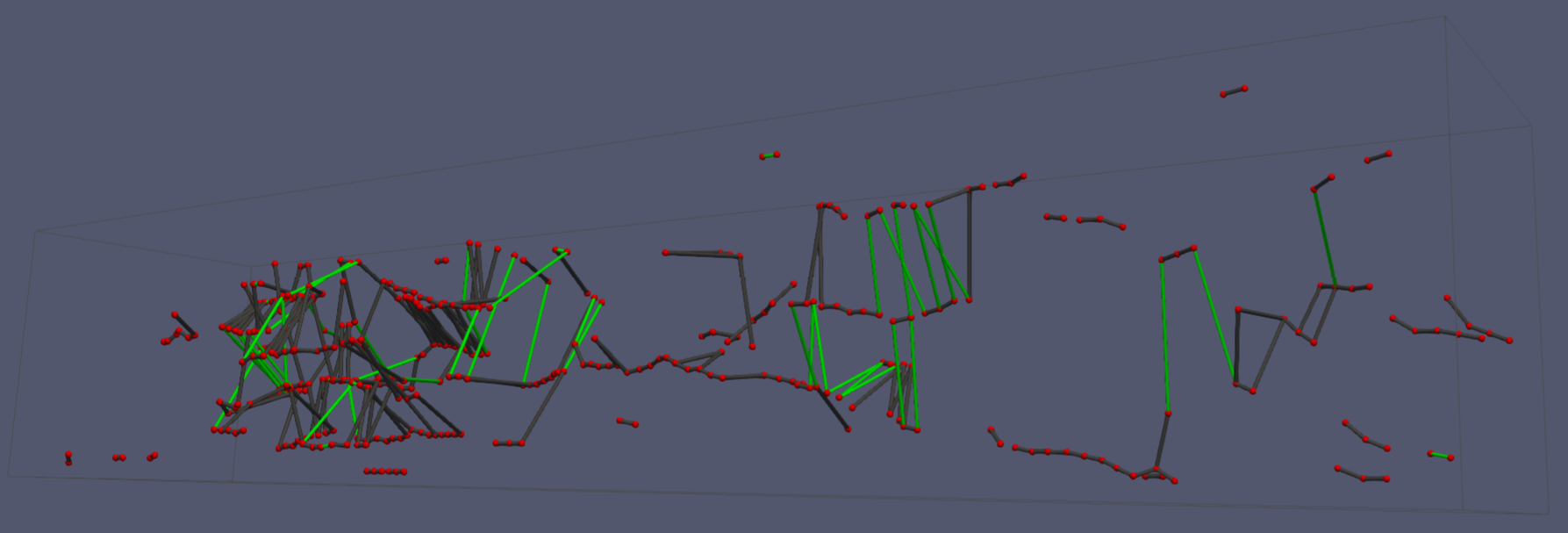}
\label{fig_app_scw_3}}
\caption{\highlightadd{Comparison of \TVEG tracks computed using different weight combinations assigned to the score components. (G,L)= (a)~(0.25,0.75), (b)~(0.5,0.5), and (c)~(0.75,0.25). The gray arcs are reported for all weight combinations. Additional arcs for a specific weight combination are highlighted in pink, yellow, and green, respectively.}}
\label{fig_app_scw_comp}
\end{figure*}

\highlightadd{In this section, we present the results of an additional study that helps understand the contributions of global and local components. This study supplements the parameter study described in the paper. In Figure~\ref{fig_app_scw_comp}, the \TVEG tracks computed for three different weight assignments to the score components are presented. The \TVEG tracks presented in Figure~\ref{fig_app_scw_1} -~\ref{fig_app_scw_3} correspond to a gradual increase in weights assigned to the global component persistence in comparison to the local components. While the gray \TVEG tracks are found in common across all three settings, the exclusive contributions are highlighted in pink, yellow, and green, respectively. We observe that increasing the weight of the persistence component results in an increase in the number of abrupt jumps within \TVEG tracks. Note that the weight assignment $(G,L)=(0.25,0.75)$ indicates equal assignment of weights to the four score components. This experiment further substantiates our initial choice of assigning equal weights to all score components for computing \TVEG tracks.}

\section{\highlightadd{Runtime analysis for \TVEG computation}}
\highlightadd{\begin{table}
    \centering
    \caption{Runtimes for \TVEG computation. The min, max, and average entries correspond to running times for processing individual time steps. Total refers to running time for processing all time steps.}
    \label{table_runtime}
    \begin{tabular}{r|l|r|r}
         \textbf{Dataset} &  & \textbf{\# Extremum} & \textbf{Running}\\
         &  & \textbf{graph nodes} & \textbf{time (ms)}\\         \hline 
         & & &  \\
         \textbf{Gauss8} & \textit{Min} & 14 & 0.21 \\
         & \textit{Avg} & 20 & 0.92 \\%20.2
         & \textit{Max} & 32 & 3.31 \\
         & \textit{Total} & 1010 & 53.69 \\
         & & &  \\
         \textbf{Viscous fingers} & \textit{Min} & 110 & 7.06\\
         & \textit{Avg} & 123 & 24.62 \\%123.32
         & \textit{Max} & 133 & 56.17 \\
         & \textit{Total} & 14798 & 2963.96 \\
         & & &  \\
         \textbf{Vortex street} & \textit{Min} & 10 & 0.30\\
         & \textit{Avg} & 82 & 18.67\\%82.45
         & \textit{Max} & 294 & 265.85\\
         & \textit{Total} & 41883 & 9595.16 \\
    \end{tabular}
\end{table}}
\highlightadd{The datasets in the case studies contain different number of critical points, which is an indicator of the complexity of dynamic behavior in the data. The time for computing the \TVEG naturally depends on the number of critical points and the size of the extremum graph. Table~\ref{table_runtime} summarizes the time taken by \textsc{TemporalArcs} for computing the \TVEG. It lists the minimum / average / maximum per time step and the cumulative running time for \TVEG computation of  the three datasets. The number of nodes in the extremum graph (total number of maxima and 2-saddles) are also reported.

We observe that the \textit{Gauss8} data requires significantly less time since it is a small dataset. While the viscous fingers dataset is smaller than the vortex street dataset, it represents sufficient complexity in terms of temporal dynamics so that the average running time is comparable. The temporal dynamics of viscous fingers may be inferred from the significant difference between the minimum and maximum running time for processing one time step. Finally, the vortex street data is most complex, in terms of temporal dynamics, amongst the three datasets. The difference between the minimum and maximum running time is most significant for this dataset. Though the average running times are comparable, the vortex street data is significantly larger in size and contains more complex temporal dynamics. There is a long temporal phase where the size of the extremum graphs are small, followed by a transition to time steps where the extremum graphs are significantly large. Hence, the cumulative running times is larger for the vortex street dataset.}

\section{\minorrevision{Visual Analysis Pipeline}}
\begin{figure*}
\centering
\includegraphics[width=0.9\textwidth]{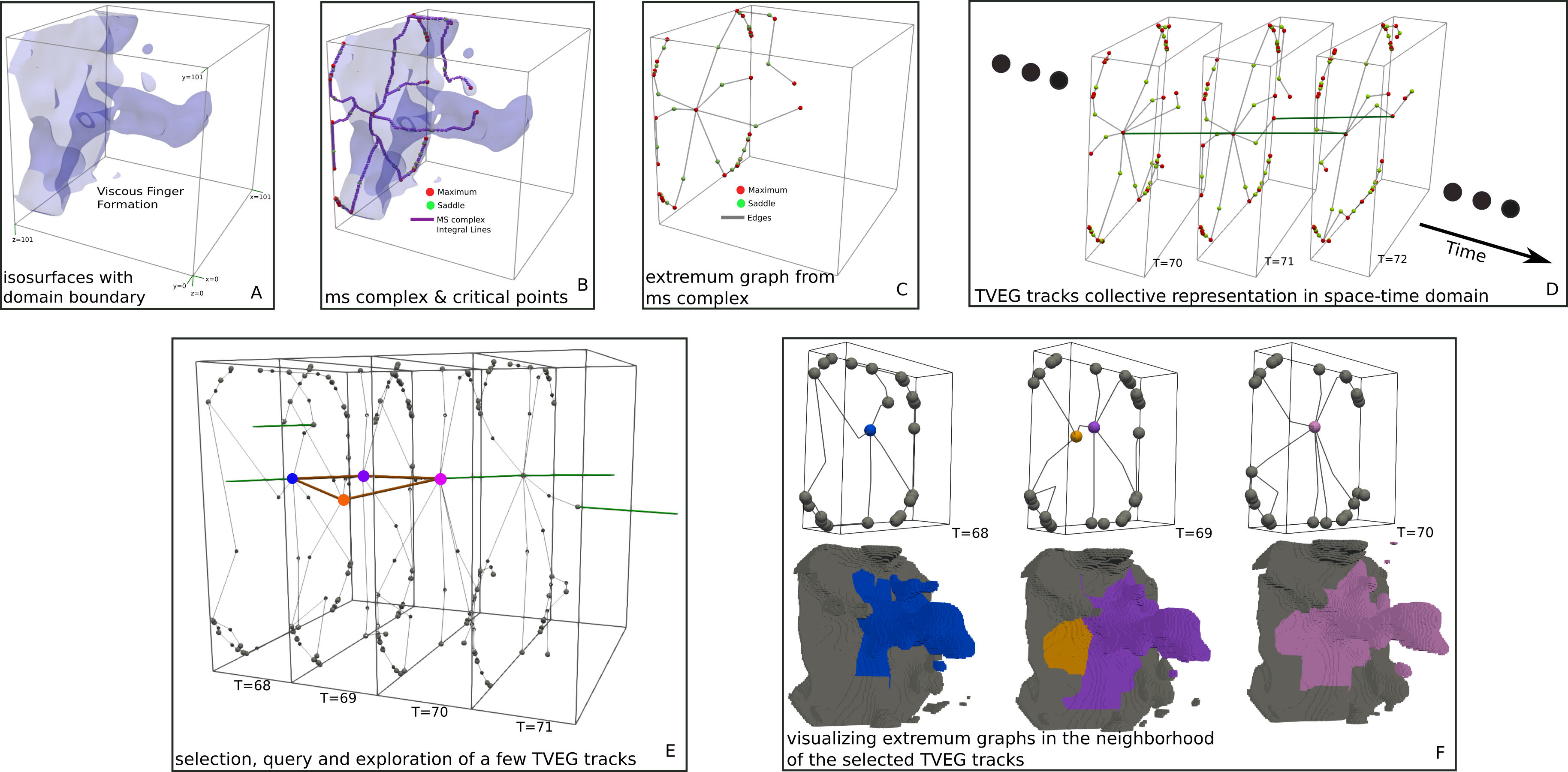}
\caption{\minorrevision{Workflow of a typical visual analysis task based on the \TVEG for the viscous fingers dataset. Individual steps include computation of the MS complex from the input scalar field~(\textbf{A}) and subsequent extraction of extremum graphs~(\textbf{B} and \textbf{C}) within each time step. Block \textbf{B} shows the integral lines representing the skeletal structure of the MS complex. \TVEG tracks shown in deep green~(\textbf{D}) are computed as a collection of arcs between a pair of extremum graphs from consecutive time steps. To avoid ambiguity between a spatial and a temporal arc, the spatial domain is scaled down along the time direction as shown in block \textbf{D}. A typical visualization task involves the selection and display of a subset of time steps~(\textbf{E}) followed by their individual inspection~(\textbf{F}). Block~\textbf{E} shows time steps that contain \TVEG tracks whose constituent maxima contribute most to the viscous finger formation. The \TVEG track of interest and the maxima in the track are highlighted~(\textbf{E}). The \TVEG supports visualization of the neighborhood (arcs of extremum graph) of individual nodes of the chosen track. It also supports visualization of the corresponding dynamic changes~(\textbf{F}) in the global data contributed by the maxima. In block~\textbf{F}, a maximum and its descending manifold are highlighted using a common color.}}
\label{fig_vispipeline}
\end{figure*}
\minorrevision{We illustrate the basic computational, user interaction, and exploration stages involved in a typical visual analysis workflow using Figure~\ref{fig_vispipeline}. Given an input time-varying scalar field, \TVEG computation begins with computation of the MS complex. The computed MS complexes at each time step  are processed to extract the extremum graph. Blocks~\textbf{B} and~\textbf{C} in Figure~\ref{fig_vispipeline} show the extremum graphs. Maxima are shown in red and 2-saddles green. \TVEG tracks are computed as a sequence of temporal arcs, which denote correspondences between maxima from consecutive time steps. Block~\textbf{D} shows portion of \TVEG tracks within three consecutive time steps. The time direction is chosen as the spatial direction of growth of the tracks (z-direction in this dataset). The domain is scaled down along the z-axis to distinguish between spatial arcs of the extremum graph and temporal arcs in the \TVEG. 

Figure~\ref{fig_vispipeline} is also used to demonstrate how \TVEG can be applied to support simultaneous visualization of data dynamics at both the extremum graph and global data level. This is possible due to the rich geometric context, in the form of the extremum graph, that is present in the \TVEG and the tracks within. Once computed, a specific portion of \TVEG tracks can be chosen to explore the temporal dynamics within the data. For instance, in Block~\textbf{E}, we select a section of \TVEG tracks highlighted in brown, where the maxima contribute significantly to finger formation dynamics. This selection includes a split followed by a merge within three consecutive time steps. This choice lets us focus on the three corresponding neighborhoods shown in Block~\textbf{F} within the extremum graphs where a maximum (in blue) at time step 68 is split into two maxima (shown in violet and orange) at time 69. The highlighted maxima at time 69 merge again into a maximum (pink) at time 70. From the extremum graph, we can extract the descending manifolds corresponding to these maxima participating in these topological events to visualize the resulting effect on the global data dynamics. In Block~\textbf{F}, we have highlighted the descending manifolds with the same color as the corresponding maximum. Thus a \TVEG can support visualization of temporal data dynamics influenced by temporal correspondence between  maxima from a chosen \TVEG track.}

\begin{table}
    \centering
    \caption{List of different fields corresponding to the critical points returned from the \textsc{ExGraph3D} subroutine (Algorithm~\ref{algo_static_eg_app})}
    \label{table_cp_fields_app}
    \begin{tabular}{r|l}
         \textbf{Fields} & \textbf{Description} \\
         \hline 
         \textit{id} & A unique id assigned to the critical point \\
         \textit{index} & Index\\ 
         $\Bar{x}$ & 3D coordinates\\ 
         $\eta$ & Neighborhood contribution \\
         \textit{pers} & Topological persistence \\
         \textit{ascmfold} & Ascending manifold\\
         \textit{dscmfold} & Descending manifold \\
         \textit{geom} & Ascending / Descending manifold geometry \\
         $t$ & The time stamp of the scalar field
    \end{tabular}
\end{table}
\begin{algorithm*}
\SetAlgoLined
\DontPrintSemicolon
\SetKwInOut{Input}{Input}\SetKwInOut{Output}{Output}
\Input{A set of extremum graphs $[\mathcal{G}^p, \ldots, \mathcal{G}^r]$}
\Output{Temporal arc set $A^{t*}$ \\ Topological event sets $\mathcal{E}^{m*}, \mathcal{E}^{s*}, \mathcal{E}^{d*}, \text{and } \mathcal{E}^{g*}$}
\BlankLine
\textbf{Initialization:} $A^{t*} \leftarrow \varnothing$; $A^t \leftarrow \varnothing$; $M^0 \gets \varnothing$; $M^1 \gets \varnothing$; \; 

\tcc{Initialize $M^0$ as maxima set of $\mathcal{G}^p$}
$M^0 \gets M^p$

\tcc{Initialize all topological event sets}
$\{\mathcal{E}^{m*}, \mathcal{E}^{s*}, \mathcal{E}^{d*}, \mathcal{E}^{g*}\} \gets\varnothing$\;
\For{$i \gets p+1\; \KwTo \; r$}{

    \tcc{Initialize $M^1$ as maxima set of $\mathcal{G}^i$}
    $M^1 \gets M^i$

    $\mathcal{S} \gets \textsc{ComputeScores}(M^0,M^1)$\;
    $\mathcal{S} \gets \textsc{FilterScores}(\mathcal{S})$\;
    \tcc{Compute the temporal arc set $A^t$}
    \ForEach{$(m^0,m^1,s) \in \mathcal{S}$}{
        $A^t\gets A^t \cup (m^0,m^1)$\;
    }
    \tcc{Detect topological events}
    $\mathcal{E}^m \gets \textsc{DetectMerge}(\mathcal{S},i)$\; 
    $\mathcal{E}^s \gets \textsc{DetectSplit}(\mathcal{S},i)$\;
    $\mathcal{E}^d \gets \textsc{DetectDel}(\mathcal{S},M^0,i)$\; 
    $\mathcal{E}^g \gets \textsc{DetectGen}(\mathcal{S},M^1,i)$\;
    \tcc{Remove z-shape configurations.}
    $\mathcal{W}\gets\mathcal{E}^m \bigcap \mathcal{E}^s$\;
    \Repeat{$\mathcal{W}=\emptyset$}
    {$w \gets \textsc{MaxScoreEdge}(\mathcal{W}$)\;
    $A^t\gets A^t \setminus w$\;
    $\mathcal{E}^m,\mathcal{E}^s  \gets\textsc{UpdateMergeSplitEdges}(\mathcal{E}^m,\mathcal{E}^s,w$)\;
    $\mathcal{W}\gets\mathcal{E}^m \bigcap \mathcal{E}^s$\;}
    \tcc{Populate temporal arc set}
    $A^{t*}\gets A^{t*}\cup A^t$\;
%    \tcc{Update conflicts in merge/split sets}
%    $\mathcal{E}^m\gets \mathcal{E}^m \setminus \mathcal{W}$, 
%    $\mathcal{E}^s\gets \mathcal{E}^s \setminus \mathcal{W}$\;
    \tcc{Update topological event sets}
    $\mathcal{E}^{m*}\gets\mathcal{E}^{m*}\cup\mathcal{E}^m$\;
    $\mathcal{E}^{s*}\gets\mathcal{E}^{s*}\cup\mathcal{E}^s$\;
    $\mathcal{E}^{d*}\gets\mathcal{E}^{d*}\cup\mathcal{E}^d$\; 
    $\mathcal{E}^{g*}\gets\mathcal{E}^{g*}\cup\mathcal{E}^g$\;
    \tcc{Re-initialize for next iteration}
    $A^t\gets\varnothing$, $M^0\gets M^1$\;
}
\caption{\textsc{TemporalArcs}}
\label{algo_tem_arcs}
\end{algorithm*}

\begin{algorithm*}
\SetAlgoLined
\DontPrintSemicolon
\SetKwInOut{Input}{Input}\SetKwInOut{Output}{Output}
\Input{Edge sets $\mathcal{E}^m$, $\mathcal{E}^s$, maximum score edge $w$ }
\Output{Modified Edge sets}
\BlankLine
\tcc{Remove edge adjacent to $w$ in $\mathcal{E}^s$ and $\mathcal{E}^m$}
$u \gets $ edge that participates in split with $w$\;
$\mathcal{E}^s\gets \mathcal{E}^s \setminus \{w,u\}$\;
$U \gets $ set of edges that participate in merge with $w$\;
\uIf{U == \{u\}}{
    $\mathcal{E}^m\gets \mathcal{E}^m \setminus u$\; 
}
$\mathcal{E}^m\gets \mathcal{E}^m \setminus w$\; 
    
\KwRet{$\mathcal{E}^m$, $\mathcal{E}^s$}\;
\caption{\textsc{UpdateMergeSplitEdges} Update the set of edges participating in merge/split events }
\label{algo_updateMergeSplitEdges}
\end{algorithm*}
\begin{algorithm*}
\SetAlgoLined
\DontPrintSemicolon
\SetKwInOut{Input}{Input}\SetKwInOut{Output}{Output}
\Input{Two maxima sets $M^0,M^1$}
\Output{A set of optimal scores $\mathcal{S} = \{(u,v,s)\} \text{ s.t.} (u,v) \in M^0 \times M^1$ and $s \in \mathbb{R}$}
\BlankLine
\textbf{Initialization: } $\mathcal{S}\gets\varnothing$\;
\ForEach{$m^0 \in M^0$}{
    $Q\gets\varnothing$\;
    \ForEach{$m^1 \in M^1$}{
        \tcc{Compute score for $(m^0,m^1)$ given weights $G, L_1, L_2, L_3$}
        
        \highlightadd{$s\gets G\lvert m^0.\text{pers} - m^1.\text{pers} \rvert + L_1\lvert \mathcal{F}(m^0.\Bar{x})-\mathcal{F}(m^1.\Bar{x})\rvert + L_2\lvert m^0.\Bar{x}-m^1.\Bar{x}\rvert_2 + L_3\lvert m^0.\eta-m^1.\eta\rvert$\;}
        $Q\gets Q \cup s$\;
    }
    \tcc{Insert two lowest scores to $\mathcal{S}$}
    $s_1\gets min(Q)$, 
    $\mathcal{S}\gets \mathcal{S}\cup (m^0,m^1,s_1)$\;
    %\tcc{Select second lowest score from $Q$}
    $Q\gets Q\setminus s_1$\;
    $s_2\gets min(Q)$, 
    $\mathcal{S}\gets \mathcal{S}\cup (m^0,m^1,s_2)$\;
}
\KwRet{$\mathcal{S}$}\;
\caption{\highlightadd{\textsc{ComputeScores} Compute all correspondences and scores for a given set of maxima}}
\label{algo_computeScores}
\end{algorithm*}
\begin{algorithm*}
\SetAlgoLined
\DontPrintSemicolon
\SetKwInOut{Input}{Input}\SetKwInOut{Output}{Output}
\Input{A list of scores $\mathcal{S}$ from Algorithm~\ref{algo_computeScores}}
\Output{A filtered version of $\mathcal{S}$}
\BlankLine
\textbf{Initialization: } $\mathcal{Y}\gets\varnothing$\;
\ForEach{$(m^0,m^1,s) \in \mathcal{S}$}{
    $\mathcal{Y}\gets\mathcal{Y}\cup s$\;
}
\tcc{Compute the mean and standard deviation of all scores}
$\mu\gets$ \textsc{mean}($\mathcal{Y}$), $\sigma\gets$ \textsc{std}($\mathcal{Y}$)\;
\tcc{Refine $\mathcal{S}$ using the threshold $\tau$}
$\tau\gets \mu + \sigma$\;

\ForEach{$(m^0,m^1,s) \in \mathcal{S}$}{
    \uIf{$s \geq \tau$}{
        $\mathcal{S}\gets\mathcal{S}\setminus (m^0,m^1,s)$\;
    }
}

\KwRet{$\mathcal{S}$}\;
\caption{\textsc{FilterScores} Filter the input set of scores based on a threshold}
\label{algo_filterscores}
\end{algorithm*}
\begin{algorithm*}
\SetAlgoLined
\DontPrintSemicolon
\SetKwInOut{Input}{Input}\SetKwInOut{Output}{Output}
\Input{A list of scores $\mathcal{S}$ from Algorithm~\ref{algo_computeScores};\\ Time step $t$}
\Output{Set of merges $\mathcal{E}^m$ in $\mathcal{S}$ between time $t$ and $t+1$}
\BlankLine
\textbf{Initialization: } $\mathcal{E}^m\gets\varnothing$\; 
\ForEach{$(m^0,m^1,s) \in \mathcal{S}$}{
    $e\gets m^1$\;
    \tcc{Count correspondences mapped to $e$}
    $c\gets 0$; $\mathcal{K}\gets\varnothing$\;
    \ForEach{$(m^0,m^1,s) \in \mathcal{S}$}{
        \uIf{$m^1 == e$}{
            $\mathcal{K}\gets\mathcal{K}\cup (m^0,t)$; $c\gets c+1$\;
        }
    }
    \tcc{Merge detected. Update $\mathcal{E}^m$}
    \uIf{$c > 1$}{
        $\mathcal{E}^m\gets\mathcal{E}^m\cup(e,t+1)\cup\mathcal{K}$\;
    }
    $c\gets 0$; $\mathcal{K}\gets\varnothing$\;
}
\KwRet{$\mathcal{E}^m$}\;
\caption{\textsc{DetectMerge} Detect merge events between two consecutive time steps}
\label{algo_detectmerge}
\end{algorithm*}
\setlength{\textfloatsep}{1pt}
\setlength{\floatsep}{1pt}
\begin{algorithm*}
\SetAlgoLined
\DontPrintSemicolon
\SetKwInOut{Input}{Input}\SetKwInOut{Output}{Output}
\Input{A list of scores $\mathcal{S}$ from Algorithm~\ref{algo_computeScores};\\ Time step $t$}
\Output{Set of splits $\mathcal{E}^s$ in $\mathcal{S}$ between time $t$ and $t+1$}
\BlankLine
\textbf{Initialization: } $\mathcal{E}^s\gets\varnothing$\; 
\ForEach{$(m^0,m^1,s) \in \mathcal{S}$}{
    $e\gets m^0$\;
    \tcc{Count correspondences mapped from $e$}
    $c\gets 0$; $\mathcal{K}\gets\varnothing$\;
    \ForEach{$(m^0,m^1,s) \in \mathcal{S}$}{
        \uIf{$m^0 == e$}{
            $\mathcal{K}\gets\mathcal{K}\cup (m^1,t+1)$; $c\gets c+1$\;
        }
    }
    \tcc{Split detected. Update $\mathcal{E}^s$}
    \uIf{$c > 1$}{
        $\mathcal{E}^s\gets\mathcal{E}^s\cup(e,t)\cup\mathcal{K}$\;
    }
    $c\gets 0$; $\mathcal{K}\gets\varnothing$\;
}
\KwRet{$\mathcal{E}^s$}\;
\caption{\textsc{DetectSplit} Detect split events between two consecutive time steps}
\label{algo_detectsplit}
\end{algorithm*}
\begin{algorithm*}
\SetAlgoLined
\DontPrintSemicolon
\SetKwInOut{Input}{Input}\SetKwInOut{Output}{Output}
\Input{A list of scores $\mathcal{S}$ from Algorithm~\ref{algo_computeScores}; \\A list of maxima $M^t$ for time step $t$; \\Time step $t$}
\Output{Set $\mathcal{E}^d$ with all the deletion events detected from  $\mathcal{S}$ between time $t$ and $t+1$}
\BlankLine
\textbf{Initialization: } $\mathcal{E}^d\gets\varnothing$; $\mathcal{K}\gets M^t$\;
\tcc{Record maxima with no edges to $t+1$}
\ForEach{$(m^0,m^1,s)\in\mathcal{S}$}{
    $\mathcal{K}\gets\mathcal{K}\setminus m^0$\;
}
\tcc{Add time step information}
\ForEach{$e\in\mathcal{K}$}{
    $\mathcal{E}^d\gets\mathcal{E}^d\cup(e,t)$
}
\KwRet{$\mathcal{E}^d$}\;
\caption{\textsc{DetectDel} Detect deletion events between two consecutive time steps}
\label{algo_detectdel}
\end{algorithm*}
\begin{algorithm*}
\SetAlgoLined
\DontPrintSemicolon
\SetKwInOut{Input}{Input}\SetKwInOut{Output}{Output}
\Input{A list of scores $\mathcal{S}$ from Algorithm~\ref{algo_computeScores}; \\A list of maxima $M^{t+1}$ for time step $t+1$; \\Time step $t$}
\Output{Set $\mathcal{E}^g$ with all the generation events detected from  $\mathcal{S}$ between time $t$ and $t+1$}
\BlankLine
\textbf{Initialization: } $\mathcal{E}^g\gets\varnothing$; $\mathcal{K}\gets M^{t+1}$\;
\tcc{Record maxima with no edges from $t$}
\ForEach{$(m^0,m^1,s)\in\mathcal{S}$}{
    $\mathcal{K}\gets\mathcal{K}\setminus m^1$\;
}
\tcc{Add time step information}
\ForEach{$e\in\mathcal{K}$}{
    $\mathcal{E}^g\gets\mathcal{E}^g\cup(e,t+1)$
}
\KwRet{$\mathcal{E}^g$}\;
\caption{\textsc{DetectGen} Detect generation events between two consecutive time steps}
\label{algo_detectgen}
\end{algorithm*}

\begin{algorithm*}
\SetAlgoLined
\DontPrintSemicolon
\SetKwInOut{Input}{Input}\SetKwInOut{Output}{Output}
\Input{A scalar field $\mathcal{F}(t)$ at time step $t$\\A persistence threshold \minoredits{$\theta$}}
\Output{Extremum Graph $\mathcal{G}^t=(V^t,E^t)$}
\BlankLine
\textbf{Initialization:} $V^t \leftarrow \varnothing$;$E^t \leftarrow \varnothing$ \;
\textbf{Compute MS complex:} $\mathcal{C} \leftarrow \textsc{MS3D}(\mathcal{F}(t),$ \minoredits{$\theta$}$)$ \;
\tcc{Store neighborhood maxima and saddles}
\ForEach {$c \in \mathcal{C}$}{\label{algo_ex_start}
    \uIf{$c.\text{index} == 2$}{ 
        $M \leftarrow c.\text{ascmfold}$ \;
        \ForEach{$m \in M$}{
            $m^t \leftarrow m$, $V^t \leftarrow V^t \cup m^t$\;
            $E^t \leftarrow E^t \cup (c.\text{id},m.\text{id})$\;
        }
    }
    $c^t \leftarrow c$, $V^t \leftarrow V^t \cup c^t$\;
}\label{algo_ex_end}
$V^t \leftarrow \text{set}(V^t)$\;
\KwRet{$\mathcal{G}^t=(V^t,E^t)$}\;
\caption{\textsc{ExGraph3D} Compute extremum graph}
\label{algo_static_eg_app}
\end{algorithm*}